\documentclass[useAMS,usenatbib]{mn2e}

\usepackage{times}
\usepackage{graphicx}
\usepackage{amssymb}
\usepackage{amsmath}
\usepackage{subfig}
\usepackage{url}
\usepackage{xspace}

\newcommand{\Msol}{M_{\odot}}
\newcommand{\Lya}{Ly$\alpha$\xspace}
\newcommand{\xa}{$x_{\mathrm{Ly}\alpha}$\xspace}
\newcommand{\xm}{\langle x \rangle_\mathrm{m} }
\newcommand{\Ctwo}{\textsc{C}$^2$-\textsc{Ray}\xspace}
\newcommand{\CPPPM}{\textsc{CubeP}$^3$\textsc{M}\xspace}
\newcommand{\Mhalo}{M_{\mathrm{h}}}
\newcommand{\ud}{\mathrm{d} }
\newcommand\ion[2]{#1$\;${\scshape{#2}}}%
\newcommand\hi{\ion{H}{i}\xspace}
\newcommand\hii{\ion{H}{ii}\xspace}

\title[Probing reionization with \Lya emitters]{On the use of \Lya emitters as probes of reionization}
\author[H. Jensen et al.]{Hannes Jensen$^{1}$\thanks{hannes.jensen@astro.su.se},
Peter Laursen$^{1,2}$, Garrelt Mellema$^{1}$, Ilian T. Iliev$^{3}$, \newauthor Jesper Sommer-Larsen$^{2,4}$, and Paul R. Shapiro$^{5}$ \\
$^{1}$Dept.\ of Astronomy and Oskar Klein Centre, Stockholm University, AlbaNova, SE-10691 Stockholm, Sweden\\
$^{2}$Dark Cosmology Centre, Niels Bohr Institute, University of Copenhagen, Juliane Maries Vej 30, DK-2100, Copenhagen \O, Denmark \\
$^{3}$Astronomy Centre, Department of Physics \& Astronomy, Pevensey II Building, University of Sussex, Falmer, Brighton BN1 9QH, United Kingdom\\
$^{4}$Marie Kruses Skole, Stavnholtvej 29-31, DK-3520 Farum, Denmark\\
$^{5}$Department of Astronomy, University of Texas, Austin, TX 78712-1083, USA}
\begin{document}

\date{\today}

\pagerange{\pageref{firstpage}--\pageref{lastpage}} \pubyear{2012}

\maketitle

\label{firstpage}

\begin{abstract}
We use numerical simulations to study the effects of the patchiness of a partly reionized intergalactic medium (IGM) on the observability of \Lya emitters (LAEs) at high redshifts ($z \gtrsim 6$). We present a new model that divides the \Lya radiative transfer into a (circum-)galactic and an extragalactic (IGM) part, and investigate how the choice of intrinsic line model affects the IGM transmission results. We use our model to study the impact of neutral hydrogen on statistical observables such as the \Lya restframe equivalent width (REW) distribution, the LAE luminosity function and the two-point correlation function. We find that if the observed changes in LAE luminosity functions and equivalent width distributions between $z \sim 6$ and $z\sim 7$ are to be explained by an increased IGM neutral fraction alone, we require an extremely late and rapid reionization scenario, where the Universe was $\sim$~40~\% ionized at $z=7$, $\sim$~50~\% ionized at $z=6.5$ and $\sim100~\%$ ionized at $z=6$. This is in conflict with other observations, suggesting that intrinsic LAE evolution at $z\gtrsim6$ cannot be completely neglected. We show how the two-point correlation function can provide more robust constraints once future observations obtain larger LAE samples, and provide predictions for the sample sizes needed to tell different reionization scenarios apart.\end{abstract}

\begin{keywords}
  galaxies:high-redshift---
galaxies:statistics --- radiative transfer --- methods: numerical
\end{keywords}

\section{Introduction}

The epoch of reionization (EoR) is currently one of the major frontiers in observational cosmology. It marks the last big phase transition of the Universe, when the intergalactic medium (IGM) went from neutral to ionized. It is generally believed (e.g. \citealt{robertson2010}) that the first stars and galaxies played a dominant role in ionizing the IGM, but the exact nature and timing of this important event in the history of the Universe have so far remained elusive.

Current observations of the EoR are all indirect, and do not provide very strong constraints. Spectra from high-redshift quasars show absorption bluewards of the \Lya wavelength --- so-called Gunn-Peterson troughs \citep{gunnpeterson65} --- implying that the Universe did not completely reionize until $z\approx6$ \citep{fan2006}. Measurements of the Thompson scattering optical depth from the WMAP satellite imply a reionization redshift of $z_{\mathbf{re}} \approx 10$ \citep{komatsu2009}, but are consistent also with various extended reionization scenarios. Measurements of the IGM temperature suggest that reionization finished no earlier than $z\sim10$ and no later than $z\sim6.5$ \citep{theuns2002,hui2003,raskutti2012}. In the future, direct constraints on the EoR are expected to come from observations of the redshifted 21~cm line emitted by the (partly) neutral IGM. Radio telescopes such as LOFAR (LOw Frequency ARray\footnote{\url{http://www.lofar.org}}), PAPER (Precision Array to Probe Epoch of Reionization\footnote{\url{http://astro.berkeley.edu/~dbacker/eor}}), 21CMA (21 cm Array\footnote{\url{http://21cma.bao.ac.cn}}) and MWA (Murchison Widefield Array\footnote{\url{http://www.mwatelescope.org}}) are just beginning observations that will measure statistical properties of the 21~cm radiation, and the planned SKA (Square Kilometer Array\footnote{\url{http://www.skatelescope.org}}) will be able to image the process directly. 

Meanwhile, galaxies with \Lya emission have started attracting attention as an additional indirect probe of the later stages of the EoR. The resonant nature of the \Lya line makes it sensitive to even small amounts of neutral hydrogen, and we thus expect the observed properties of \Lya emitters (LAEs) to change at higher redshifts, when the IGM was on average more neutral.

In the last few years, there have been many efforts to obtain large high-$z$ samples of LAEs (e.g.\ \citealt{malhotrarhoads2004,ouchi2010, hu2010,kashikawa2011}), using narrow-band photometric observations and in some cases with spectroscopic confirmations. We are now starting to get sizable samples at redshifts where reionization might still be ongoing. So far however, attempts to obtain constraints on the global IGM neutral fraction from these samples have produced somewhat different and contradictory results.

There are a number of ways one could potentially study the EoR using LAEs. Some studies (\citealt{dayal2011,dijkstra2011}) try to constrain the ionization state of the IGM based on individual detections of very high-$z$ LAEs \citep{lehnert2010}. The reasoning is that the mere observation of a single LAE would set an upper limit on the amount of neutral hydrogen present around the galaxy --- were it too high, the galaxy would not have been observable. 

A more common approach, and the one used in this paper, is to focus instead on the effects that an increasingly neutral IGM at higher redshifts will have on the \emph{statistical} properties of the population of LAEs. Perhaps the most obvious effect is that the increased \ion{H}{i} density at higher redshifts is expected to lead to a decrease in the fraction \xa of UV-selected galaxies that also show \Lya emission \citep{kashikawa2006}. While the observed LAE fraction increases with redshift up to $z\approx6$ --- likely due largely to the decreasing dust content in earlier galaxies (e.g.\ \citealt{stark2010})--- there are some tantalising observational indications that \xa might start to decrease for even higher $z$ \citep{ota2008,stark2010,pentericci2011,ono2011,schenker2012}. \cite{hayes2011} see a corresponding sharp drop in the global \Lya escape fraction. However, the uncertainties are large, and not everyone sees this effect (e.g.\ \citealt{tilvi2010}).

An observable related to the LAE fraction is the LAE luminosity function (LF). Observations show very little evolution in the LAE LF up to $z \approx 6$ (\citealt{malhotrarhoads2004,ouchi2008,tilvi2010}). If \Lya radiation becomes attenuated by an increasingly neutral IGM at higher redshifts, we expect the LF to start dropping in amplitude. There are indeed some observations that indicate this (\citealt{ouchi2010,kashikawa2011}), although the data remains limited, especially at the faint and bright ends of the LF. If there is a connection between the \Lya luminosity of a galaxy and the transmitted fraction of \Lya photons, this could also change the shape of the LF. 

A third potentially useful observable is the two-point correlation function of LAEs, which measures the clustering of these galaxies in the sky. The patchiness of the reionization process is expected to give rise to an increase in the clustering of galaxies (\citealt{mcquinn2007,iliev2008a,tilvi2009,ouchi2010}). Since the \Lya transmission depends on the ionization state of the IGM in the vicinity of a galaxy, observations will favour galaxies located within large \hii regions. Thus, the two-point correlation function will show a stronger clustering for a sample of LAEs than for a UV-selected sample with the same number density.

Accurate numerical modelling of observed LAE populations at very high redshifts is a computationally challenging task. It involves simulating the growth of structure in the early Universe, the patchy reionization of the IGM (where many essential parameters are still poorly constrained), star formation in the first galaxies, and finally the complicated radiative transfer (RT) of \Lya photons. Most previous studies use simplified semi-analytical models of the IGM (\citealt{dayal2008,tilvi2009,dijkstra2011}), or very simple models of the \Lya emitters themselves (\citealt{mcquinn2007,iliev2008a}). Others use detailed models of both the IGM and the LAEs \citep{zheng2010}, which comes at the cost of a higher computational complexity, resulting in a smaller part of parameter space being explored.

In this paper, we use numerical models of the IGM combined with a simple recipe for \Lya line shapes --- based on detailed hydrodynamical simulations of galaxies --- to obtain predictions for the statistical observables described in the introduction. Like \cite{zheng2010}, we only consider one specific reionization scenario, but by simplifying the radiative transfer of \Lya through the IGM (see Sec.~\ref{sec:igmtransfer}) we are able to cover a large range of IGM ionized fractions. This study is similar in nature to \cite{iliev2008a}, but uses an updated version of the reionization code, a bigger simulated box and a novel method for separating the complicated \Lya radiative transfer in the close vicinity of galaxies from the more straightforward calculations in the low-density IGM. Whereas \cite{iliev2008a} simply assumed a Gaussian \Lya line shape for all emitters, we use a recipe for double-peaked profiles with little emission at the line centre, motivated by spectra from detailed radiative transfer in the vicinity of galaxies taken from cosmological Smooth Particle Hydrodynamics (SPH) simulations. We compare the results from this line profile to the simple Gaussian model, and discuss some implications of various assumptions regarding the choice of line shape model.

The outline of the paper is as follows: in Section \ref{sec:simulations} we describe our simulations and the assumptions that went into the modelling. In Section \ref{sec:results} we present the results from the simulations. We discuss the line shape model used and the implications of using different line models, and the effect of the IGM on the observed line shapes. We show our simulated luminosity functions, equivalent width distributions and correlation functions and compare these to observations where available. Finally, in Section \ref{sec:discussion}, we summarise and discuss our results.

The cosmological parameters used in the simulations were for a flat $\Lambda$CDM model with $(\Omega_{m}, \Omega_b, h, n, \sigma_8) = (0.27, 0.044, 0.7, 0.96, 0.8)$, consistent with the 5 year WMAP results \citep{komatsu2009}.


\section{Simulations}
\label{sec:simulations}
\begin{figure*}
	\includegraphics[width=14cm]{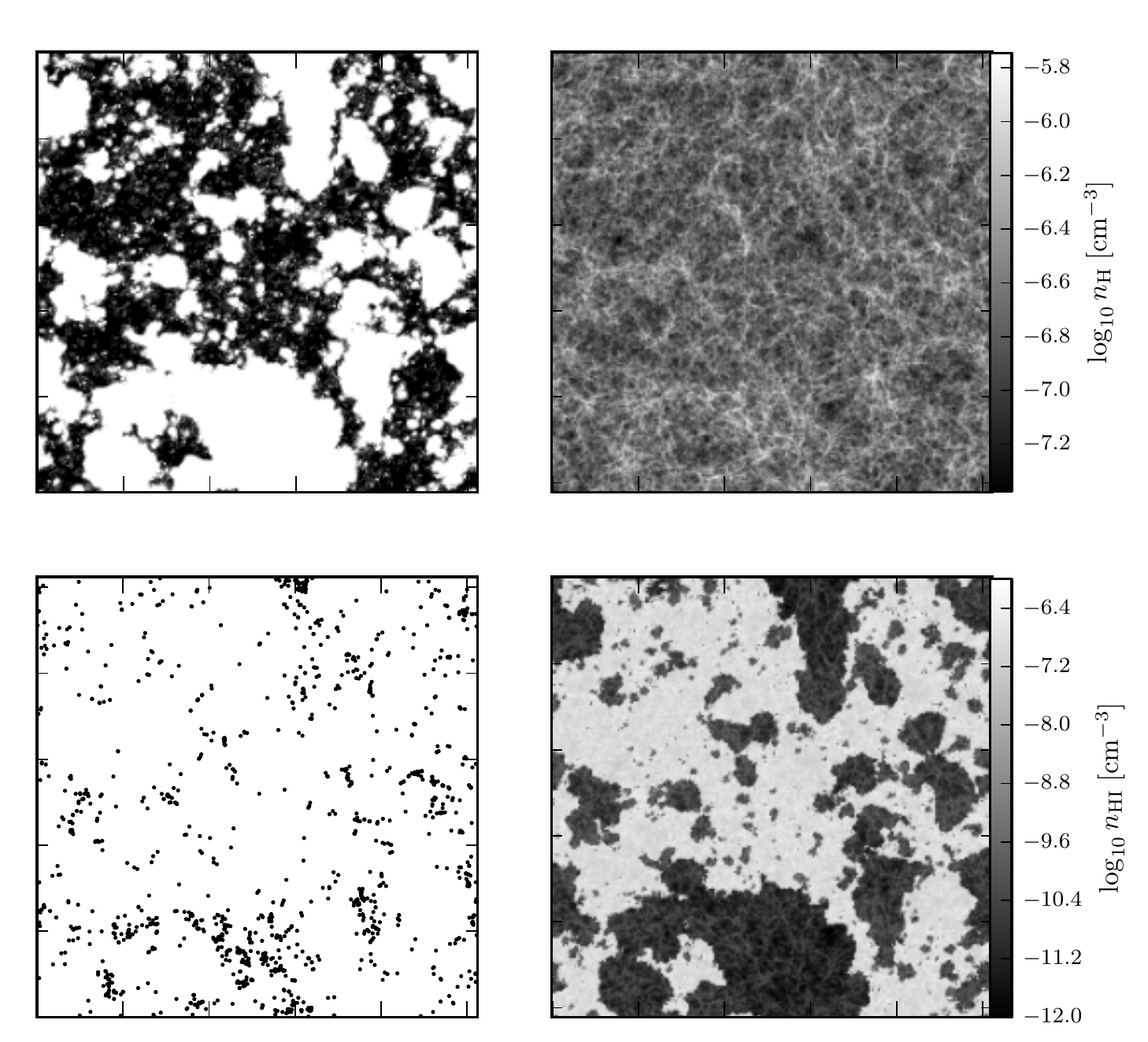}
	\caption{Output from our IGM simulations at a global ionized fraction $\xm= 0.67$. The figures show a 1-cell thick ($=0.64$ cMpc) slice of the IGM. All figures are $163$~cMpc across. \emph{Top left}: Ionized fraction (white is ionized, black is neutral). \emph{Top right}: \hi+\hii density. \emph{Bottom left}: Halo positions for $\Mhalo > 10^{10} \Msol$. \emph{Bottom right}: Neutral hydrogen density.}
	\label{fig:ionization_slice}
\end{figure*}

Our simulations were carried out in four separate steps. First, a cosmological N-body simulation was performed to obtain a time evolving density field and dark matter halo lists. The density field was then used to simulate the transfer of ionizing radiation and the growth of the \hii regions surrounding the haloes. Separately from this, a number of galaxies were simulated in detail with a hydrodynamical code, modelling the complicated \Lya radiative transfer in the close vicinity of the galaxies. We used the spectra derived from the hydrodynamical simulations to get a picture of what the \Lya line looks like at the edge of the galaxies, and how it varies in shape with halo mass in order to devise a simplified line model. Finally, after assigning \Lya spectra to the dark matter haloes from our N-body simulations, we calculated the optical depth, $\tau(\lambda)$, through the IGM by tracing a large number of sightlines from each galaxy through our simulation box, taking into account the variations in density, ionized fraction and gas peculiar velocities.

\subsection{N-body simulations}
The N-body structure formation simulations were performed with the massively-parallel, hybrid (\textsc{MPI}+\textsc{OpenMP}) code \CPPPM (\citealt{iliev2008b,harnoisderaps2012}), developed from the \textsc{PMFAST} code \citep{merz2005}. \CPPPM is a particle-particle-particle-mesh code which calculates gravitational force interactions by combining long-range forces calculated on a mesh with short-range direct forces between particles. For efficiency reasons the long-range forces are calculated on two grids: a fine grid covering the local domain, and a global coarse grid. 

A spherical over-density method was used to identify dark matter haloes. This method starts by identifying local maxima in the fine grid density distribution, and proceeds by enclosing these in successively larger spheres until the average density inside the sphere is below $178$ times the global mean density. The N-body code was run on a cube with a side of $163$ comoving Mpc (hereafter cMpc) with a fine grid of $6144^3$ points using 3072$^3$ particles, and finally regridded onto a grid with 512$^3$ cells. This box size is large enough that cosmic variance effects are small \citep{barkana2004}. The smallest resolved haloes consisted of 20 particles, which gives a minimum halo mass of $10^8 \Msol$, corresponding to the mass at which atomic cooling becomes important (see \citealt{iliev2011} for details).

\subsection{Radiative transfer of ionizing radiation}
For the simulation of the radiative transfer of ionizing radiation from the galaxies identified in the previous step, we used the code \Ctwo --- \emph{Conservative, Causal Ray-tracing method} \citep{mellemac2ray}. In \Ctwo, each source --- taken from the N-body runs described above --- is given a flux proportional to its mass, $\Mhalo$, so that the photoionization rate from each source at a distance $r$ is:
\begin{equation}
	\Gamma(r) = \frac{1}{4 \pi r^2} \int_{\nu_{\mathrm{th}}}^{\infty}g_{\gamma}\frac{\Mhalo \Omega_b}{10 \Omega_0 m_p} \sigma_{\nu}e^{-\tau_{\nu}(r)} \ud \nu \;\; [\mathrm{Myr^{-1}}],
	\label{ionizingflux}
\end{equation}
where $\sigma_{\nu}$ is the ionization cross-section for hydrogen, $\tau_{\nu}$ is the optical depth, $m_p$ is the proton mass and $\nu_{\mathrm{th}}$ is the threshold energy for ionization of hydrogen. The factor $g_{\gamma}$ is a model parameter which determines the halo mass -- ionizing flux relation. More precisely, it is the number of ionizing photons escaping the galaxy per halo baryon per 10 Myr, which depends on assumptions about the initial mass function, star formation efficiency and escape fraction of the galaxies. We assume that the baryonic density follows the dark matter density, and we do not consider contributions to the ionization by other sources, such as QSOs.

The time evolution for the ionized fraction of hydrogen, $x_i$, is:
\begin{equation}
	\frac{\ud x_i}{\ud t} = (1-x_i)(\Gamma + n_e C_H) - x_i n_e \alpha_H,
	\label{dxdt}
\end{equation}
where $n_e$ is the electron density, $C_H$ is the collisional ionization coefficient and $\alpha_H$ is the recombination coefficient. The on-the-spot approximation is used, i.e.\ all recombinations to the ground state are assumed to be locally absorbed, so that $\alpha_H = \alpha_B$. Eq.~\eqref{dxdt} is solved by \Ctwo through an iterative process, which is necessary since $\Gamma$ can have contributions from many sources. This iteration is carried out for all sources in a photon-conserving manner, over a $(163\; \mathrm{cMpc})^3$ box, discretised into a grid with $256^3$ cells, using the density field from the N-body simulations as input. In the end, we obtain density, gas peculiar velocity, ionized fraction and ionizing flux for each cell (see Fig.\ \ref{fig:ionization_slice}). For a detailed description of \Ctwo along with several comparisons against exact analytical solutions and other radiative transfer codes, see \cite{mellemac2ray} and \cite{iliev2006}.

\begin{figure}
	\centering
	\includegraphics[width=\columnwidth]{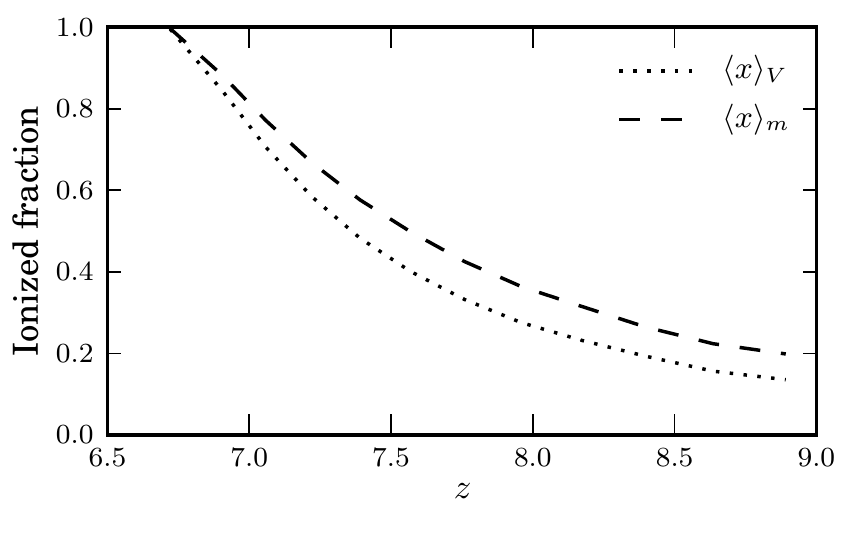}
	\caption{Evolution of the ionization state of the IGM during the late stages of our simulations. The dotted line shows the mean ionized fraction by volume, and the dashed line shows the mean ionized fraction by mass.}
	\label{fig:ionization_hist}
\end{figure}

In this study, we consider a model where $g_{\gamma}=1.7$ and where small sources ($\Mhalo < 10^{9} \Msol$) become suppressed as soon as the IGM in their vicinity is $\geq 10\%$ ionized. This effect is known as \emph{self-regulation} and is motivated by the fact that smaller sources lack the gravitational well needed to keep accreting material in an ionized and heated environment, and so will essentially stop forming stars as their neighbourhood becomes ionized \citep{iliev2007}. The $g_{\gamma} = 1.7$ model gives a fairly late reionization, consistent with observations of quasar spectra. The evolution of the global ionized fraction for hydrogen is illustrated in Fig.~\ref{fig:ionization_hist}, which shows the global ionized fraction averaged by mass and by volume. The mass-weighted ionized fraction, $\xm$ is higher than the volume-weighted fraction, indicating that the high-mass regions with many galaxies are the ones to become ionized first. This is known as \emph{inside-out} reionization. 

Due to this finite resolution, small scale density fluctuations in the
IGM are smeared out. Since recombinations scale with
$n_\mathrm{HII}n_e$, the effects of these density fluctuations are
often included by boosting the recombination term with a clumping
factor, $C=\langle n^2 \rangle/\langle n \rangle^2$. The value of this
term depends on baryonic physics and will be different between cold
and photo-heated gas. Including a maximal clumping factor based on
dark matter clumping down to scales $\sim 10^5 \Msol$ is found to extend reionization by $\Delta z\sim
1$ (Mao et al., in prep.). Lyman limit systems are the denser
structures which contain enough \hi to achieve an optical
depth $>1$. They will limit the distance ionizing radiation can
travel. Unfortunately, the evolution of this mean free path during the
EoR is not known, but their presence is likely to delay the end of
reionization by making it more difficult for ionized regions larger
than the mean free path to merge with others.

The simulation used here does not include the effects of gas clumping and
Lyman limit systems and therefore likely overestimates the redshift of
the end of reionization. On the other hand there is considerable uncertainty
in the luminosity of ionizing radiation escaping from galaxies and more
efficient sources would reionize the Universe earlier. Given these
uncertainties, we will refer to our simulated IGM boxes by mass-weighted ionized fraction, $\xm$, rather than redshift. Effectively, we are assuming that the curve in Fig.\ \ref{fig:ionization_hist} could in reality be translated slightly along the $x$-axis, i.e.\ reionization could be taking place somewhat earlier or later than in our simulations. We believe that this is reasonable if the actual topology of reionization is not substantially different from our model \citep{friedrich2011,iliev2011}.

\subsection{\Lya radiative transfer through the IGM}
\label{sec:igmtransfer}
\begin{figure}
	\centering
	\includegraphics[width=\columnwidth]{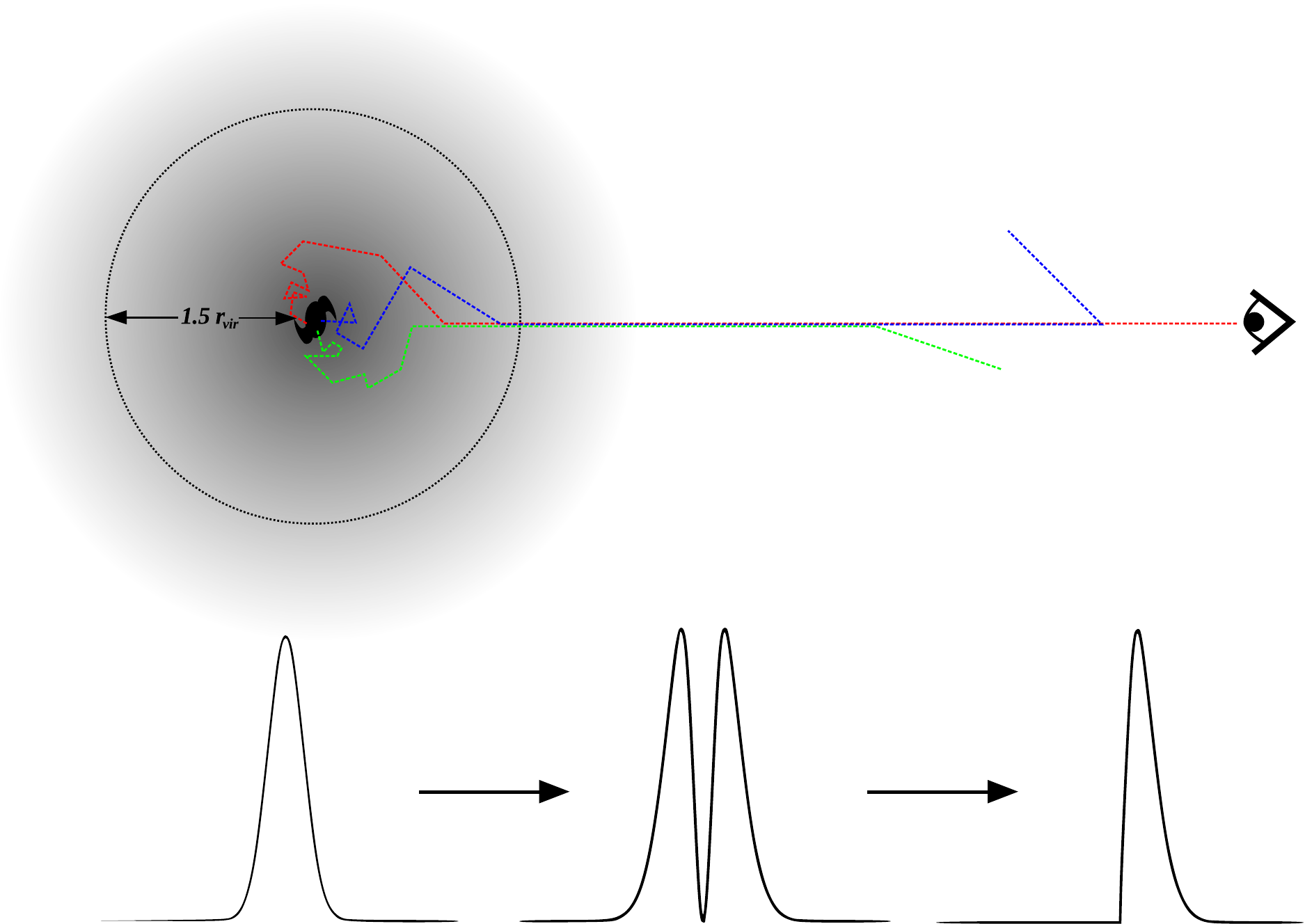}
	\caption{The two different radiative transfer regimes in our model. In the dense medium inside and close to the halo, the line --- initially with a Voigt profile --- can become broadened and shifted and typically obtains an absorption feature at the \Lya line centre. In this optically thick regime, photons scatter both into and out of the line of sight. As the radiation enters the less dense IGM, scatterings into the line of sight become extremely rare, and it becomes justified to consider only scatterings out of the sightline. Figure adapted from \protect\cite{laursen2010}. Note that the colours are only for telling different photons apart; they are not related to photon wavelengths.}
	\label{fig:line_evolution}
\end{figure}

While the general problem of \Lya radiative transfer is complex and computationally demanding (e.g.\ \citealt{zheng2010}), it can be greatly simplified in the low-density IGM far away from the galaxy where the radiation was emitted. In the high-density region close to the galaxy, photons are scattered in and out of the line of sight, undergoing frequency changes on the way. Simulating this process requires very high resolution and many assumptions regarding, for instance, dust content and star formation. However, as was shown in \cite{laursen2011}, after a distance of $\sim1.5$ times the virial radius of the galaxy, very few photons are scattered into the line of sight, and therefore it is justified to divide the radiative transfer into two regimes: a galaxy part where photons diffuse out of the optically thick gas around the halo (discussed in Sec.~\ref{sec:linemodel}), and an IGM part where we consider only scatterings \emph{out} of the line of sight (effectively absorption). We set the boundary between these two regimes to be at $1.5\; r_{\mathrm{vir}}$. This division is schematically outlined in Fig.\ \ref{fig:line_evolution}.

The (circum-)galactic part will be discussed in the next section. For the extragalactic part, we use a modified version of the code \textsc{IGMtransfer} (\citealt{laursen2011}). \textsc{IGMtransfer} calculates the \Lya transmission function $F(\lambda)$ for a given wavelength interval, i.e.\ the transmitted fraction of radiation as a function of wavelength. Note that this is independent of the shape of the emission line.

The transmission function is given by:
\begin{equation}
	F(\lambda) = e^{-\tau(\lambda)},
	\label{eq:transmission}
\end{equation}
where the optical depth, $\tau$, is calculated by stepping through the box along a line of sight, while summing the $\tau$ from each cell:
\begin{equation}
	\tau(\lambda) = \sum_i^{\mathrm{cells}} n_{\mathrm{HI,i}} \sigma_{\mathrm{Ly}\alpha}(\lambda + \lambda v_{||,i}/c) \Delta r.
	\label{eq:optical_depth}
\end{equation}
Here, $\Delta r$ is the length of each step, $n_{\mathrm{HI,i}}$ is the neutral hydrogen density, $\sigma_{\mathrm{Ly}\alpha}$ is the neutral hydrogen cross section to \Lya scattering and $v_{||,i}$ is the gas velocity component of the cell along the line of sight, including the Hubble flow. The sightline is traced until photons at $1190\;$\AA~ --- well outside the \Lya line profiles considered here --- have redshifted into resonance. The IGM temperature was assumed to be $10^4$ K everywhere.

In the original version of \textsc{IGMtransfer}, photons are emitted in random directions and bounce randomly as they hit the edge of the box. Since our simulated box uses periodic boundary conditions, we modified the code so that sightlines travel through the periodic box instead of bouncing. We also added the possibility to more finely sample the IGM box. While the original version of \textsc{IGMtransfer} is limited to sampling each cell only once, we now divide each cell into many smaller subdivisions ($\Delta r$ in Eq.\ \ref{eq:optical_depth}), which gives more reliable results for $F(\lambda)$\footnote{While we still assume $n_{\mathrm{HI}}$ to be the same across the cell, the Hubble flow velocity can vary considerably across a cell for coarser resolutions. If the cell is only sampled once, this will result in an unphysical shape of  $\tau(\lambda)$.}. Furthermore, we added the possibility to specify sightline directions, rather than having photons emitted in random directions.



A cosmological simulation unavoidably involves a trade-off between resolution
and volume. 
\citet{bolton2009} studied the impact of resolution and box size on various
observables in cosmological hydro-simulations at $2\lesssim z\lesssim5$ and
found that a gas particle mass resolution of $M \sim 10^{5\textrm{--}6}
M_\odot$ is needed for the simulations to converge. However, at the high
redshifts that we have been studying, the neutral fraction in the
lowest-density regions, i.e. far from the galaxies, first of all is still
sufficiently high that virtually no radiation is transmitted, and second
corresponds to wavelengths in the transmitted spectra very far from the
\Lya line centre. Conversely, the regions that \emph{do} affect the line are so close to
the galaxies that they typically are much better resolved in the underlying N-body simulations.

In general, a lower resolution will smear out the underlying density field and give a higher optical
depth, as seen in \cite{bolton2009}. However, a higher resolution may also start to resolve very dense, self-shielded objects that give rise to damping wings and actually lower the transmission, as shown recently by \cite{bolton2012}. Both of these studies use a spatially uniform ultraviolet background, and it remains to be seen exactly how the results will be affected by a more realistic, fluctuating ionizing flux.

%


\subsection{\Lya line model}
\label{sec:linemodel}
From the previous steps we have snapshots of the IGM and dark matter density, gas peculiar velocities, halo lists and a method for calculating the transmission of \Lya in the low-density IGM regime. The only missing piece of the puzzle is the emitted spectra $J_{\mathrm{em}}(\lambda)$, i.e.\ the \Lya spectra as they look as they emerge from the galaxies, here taken to be at a distance $1.5\;r_{\mathrm{vir}}$ from the galaxy centre, as discussed earlier. These are to be multiplied by $F(\lambda)$ to get the observed spectra. 

Some previous studies (\citealt{dijkstra2007,mcquinn2007,iliev2008a}) simply assume a Gaussian line shape, and apply Eq.\ \eqref{eq:transmission} directly to this. While this may give a rough picture of the effects of the IGM, it ignores the fact that Eq.\ \eqref{eq:transmission} is only valid in the very low density IGM. For the (circum-)galactic part (within $1.5\; r_{\mathrm{vir}}$), full radiative transfer is required, along with a detailed model of the structure of the galaxies. 

We will use the term \emph{intrinsic line} to mean the \Lya line at the border between the ISM and the IGM. For our fiducial line model (introduced below), this is taken to be $1.5\;r_{\mathrm{vir}}$. However, when we compare to the Gaussian line results, we will start the sightlines at the halo centres, essentially treating the halo as part of the IGM. This is done to facilitate fair comparisons to earlier studies, such as \cite{iliev2008a}.

\begin{figure}
	\centering
	\includegraphics[width=\columnwidth]{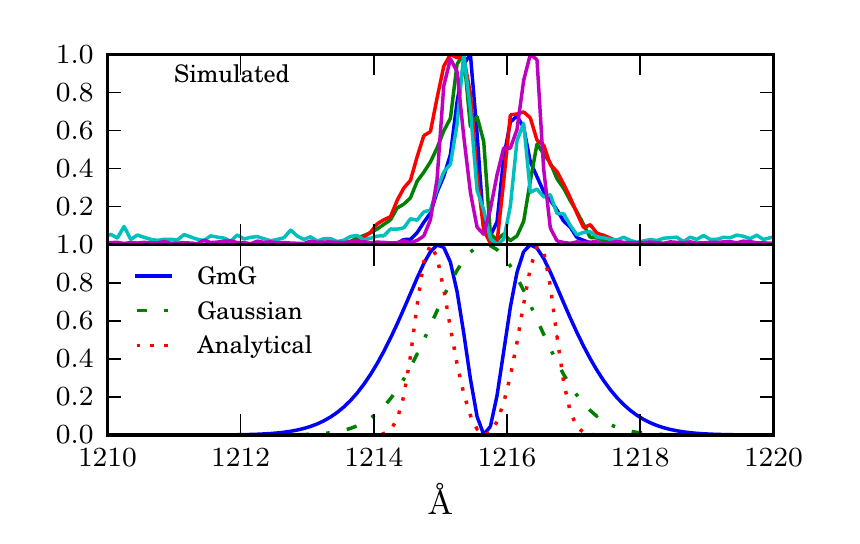}
	\caption{Some examples of simulated intrinsic \Lya line profiles (top), and the various simplified models discussed in the text (bottom). All line profiles are normalised with arbitrary units on the $y$-axis.}
	\label{fig:example_lines}
\end{figure}

In the case of a spherically symmetric, homogeneous \hi distribution, the problem of \Lya radiative transfer can be solved analytically \citep{dijkstra2006}. The resulting line shape emerging from the galaxy has the form:
\begin{equation}
	J_{\mathrm{em}}(x) = \frac{\sqrt{\pi}}{\sqrt{24}a\tau_0}\left( \frac{x^2}{1 + \cosh(\sqrt{2\pi^3/27}|x^3|/a\tau_0)} \right)
	\label{analytical_line}
\end{equation}
where $x$ is the frequency shifted to the \Lya line centre and divided by the Doppler width of the line, $a$ is the Voigt parameter and $\tau_0$ is the optical depth to the surface of the galaxy. This line is plotted in Fig.\ \ref{fig:example_lines} for some typical galaxy parameters (dotted red line in the bottom part of the figure).

Observed \Lya lines will deviate significantly from Eq.\ \eqref{analytical_line} since real galaxies are neither homogeneous nor spherical but disk-like and clumpy and may have complicated outflows and inflows. To investigate how these factors complicate the spectra, we isolated a smaller, high-resolution, sample
of $\sim 2000$ galaxies from a separate cosmological SPH simulation at four
different redshift epochs ($z = [6.0, 7.0, 7.8, 8.8]$) and ran detailed \Lya\
RT calculations on these.  

The cosmological simulation is similar to the ones outlined in \citet{laursen2011},
which themselves are a significant improvement on the ones described in
\citet{sommerlarsen2003} and \citet{sommerlarsen2006}. The reader is referred to these papers for a
detailed description of the code; in short, it is a TreeSPH code using
self-consistent, ab initio hydro/gravity simulations to calculate the structure
in a spherical region of linear dimension $10h^{-1}$ cMpc.
The mass of the SPH and DM particles were $7.0\times10^5$ $\Msol$ and
$3.9\times10^6$ $\Msol$, respectively, and the minimum smoothing length $\sim$~100 pc. A simplified ionizing UV RT scheme is
invoked, modelling the meta-galactic UV background (UVB) after \citet{haardt96}. 

The galaxies span several orders of magnitude in stellar mass; from $10^8$ to $10^{10}$
$\Msol$. Before the \Lya\ RT, the physical parameters of the gas particles
(neutral hydrogen density, temperature, bulk velocity, dust density, and \Lya\
emissivity from both cooling radiation and recombinations following ionization; at these high redshifts,
the two processes contribute roughly equally to the total \Lya luminosity)
are first interpolated onto a mesh of adaptively refined cells, using their
original kernel function. The resulting galaxies then each consist of
$10^{2\textrm{--}3.5}$ cells. However, although the physical size of the smallest cells
is $\sim 10$ pc, \emph{typically} the best resolution is in fact some 100~pc. Finally, the full RT is conducted using the Monte Carlo code {\sc MoCaLaTA} \citep{laursen2009,laursen2009b}, calculating the spectra
emerging at a distance 1.5 virial radii from the centre of the galaxies by tracing
individual photon packets as they scatter out through the ISM.


Simulating \Lya production and radiative transfer in very high-$z$ galaxies involves many poorly constrained parameters such as star formation rates, dust content, ISM ionization structure and gas clumpiness. While the simulations in our high-resolution galaxy sample give a fairly detailed picture of the \Lya line, the resolution is still too low to fully trust all aspects of the results as-is. For instance, we find that the values of the \Lya luminosities are somewhat too low to match observations. However, we assume that the general shapes of the emerging \Lya lines and their dependence on halo mass are reasonable.

A small number of the simulated lines are plotted in the top panel of Fig.\ \ref{fig:example_lines}. We find that the lines in general have a double-peaked structure, with close to zero emission at the line centre, in agreement with Eq.\ \eqref{analytical_line}. However, the two peaks are generally asymmetric, falling off more slowly away from the line centre. We also find that the lines tend to become broader for higher-mass galaxies. This general line shape is familiar from both observations and other simulations (e.g.\ \citealt{tapken2007,verhamme2008,yamada2012}). The presence of neutral hydrogen in the ISM makes it difficult for \Lya photons to escape before they have scattered and become red- or blueshifted away from the line centre. 

While not all observed lines show the double-peaked line profile, they typically have most of the emission at wavelengths away from the line centre. This is in contrast to the Gaussian model, where the peak of the emission is exactly at the line centre. As we shall see in Sec.\ \ref{sec:results}, this difference has a significant effect on the \Lya transmission through the IGM.

To model the line shapes from our high-resolution galaxy sample, we adopt a Gaussian-minus-a-Gaussian (GmG) line shape with a width that depends on the halo mass:
\begin{equation}
	J_{\mathrm{em}}(\lambda) \propto e^{-(\lambda-\lambda_0)^2/2\sigma_1^2}-e^{-(\lambda-\lambda_0)^2/2\sigma_2^2}.
	\label{gmg}
\end{equation}
By fitting such lines to all the spectra in our high-resolution galaxy sample, we find that $\sigma_1$ and $\sigma_2$ both increase with halo mass, roughly as:
\begin{align}
	\sigma_1 = -6.5 + 0.75 \log \Mhalo/\Msol  \label{sigma1}\\
	\sigma_2 = -3.2 + 0.35 \log \Mhalo/\Msol, \label{sigma2}
\end{align}
that is, the line shapes tend to become broader for more massive galaxies.

While somewhat \emph{ad-hoc}, the GmG model appears to reproduce the simulated line shapes rather well. As we will show in Sec.\ \ref{sec:results}, the important distinction when it comes to IGM transmission is whether most of the emission is at the line centre (as in the Gaussian model) or offset from the line centre (as in the GmG model), and not in the exact shape of the peaks. Since low-$z$ observations, analytical models and simulations all tend to show line shapes with either double peaks or a single peak on the blue side of the line centre (e.g.\ \citealt{verhamme2008,dijkstra2006,dijkstra2006partII,wilman2005}), we adopt Eqs.\ \eqref{gmg}, \eqref{sigma1} and \eqref{sigma2} as our fiducial model.

Observed \Lya line profiles --- in contrast to our model spectra --- often have one peak that is stronger than the other. Gas that is falling into the galaxy may cause the red peak to become weaker than the blue one, and outflows will have the opposite effect (e.g.\ \citealt{dijkstra2011}). Such asymmetries can have a large effect on the absolute value of the fraction of \Lya photons that are transmitted through the IGM. However, in this paper we calibrate our intrinsic luminosities against observations (see Sec.\ \ref{sec:intrinsic_luminosity}), which means that the absolute values of the transmitted fractions have little or no effect on our results. What we are interested in is mainly the dependence of the transmission on quantities such as the local \hi density and halo mass. These dependencies will be largely unaffected by asymmetries in the peaks. 

In our analyses, we generally make the assumption that the intrinsic properties of LAEs (i.e.\ luminosities and \Lya line shapes) do not change during the time interval we are studying. This is a strong assumption, and one that is widely debated. \cite{kashikawa2011} argue that it is probably reasonable at least up to $z=6.5$ given that observations of the UV luminosity function --- which is insensitive to the IGM --- stays the same between $z=5.7$ and $z=6.5$ within the error bars. Others, such as \cite{ouchi2010}, \emph{do} observe evolution in the UV LF, implying that the changes in LAE populations are due at least in part to intrinsic source evolution. In Sec.\ \ref{sec:discussion}, we discuss the implications of the no-intrinsic-evolution assumption on our results.

\section{Results}
\label{sec:results}
\subsection{General effects of the IGM}
Using the methods described above, we assign spectra according to Eqs.\ \eqref{gmg}, \eqref{sigma1} and \eqref{sigma2} to all haloes in our box with  $\Mhalo > 10^{10} \Msol$. We then trace 50 sightlines per halo in different directions through the IGM box and calculate the IGM transmission function for each one.

\begin{figure}
	\centering
	\includegraphics[width=\columnwidth]{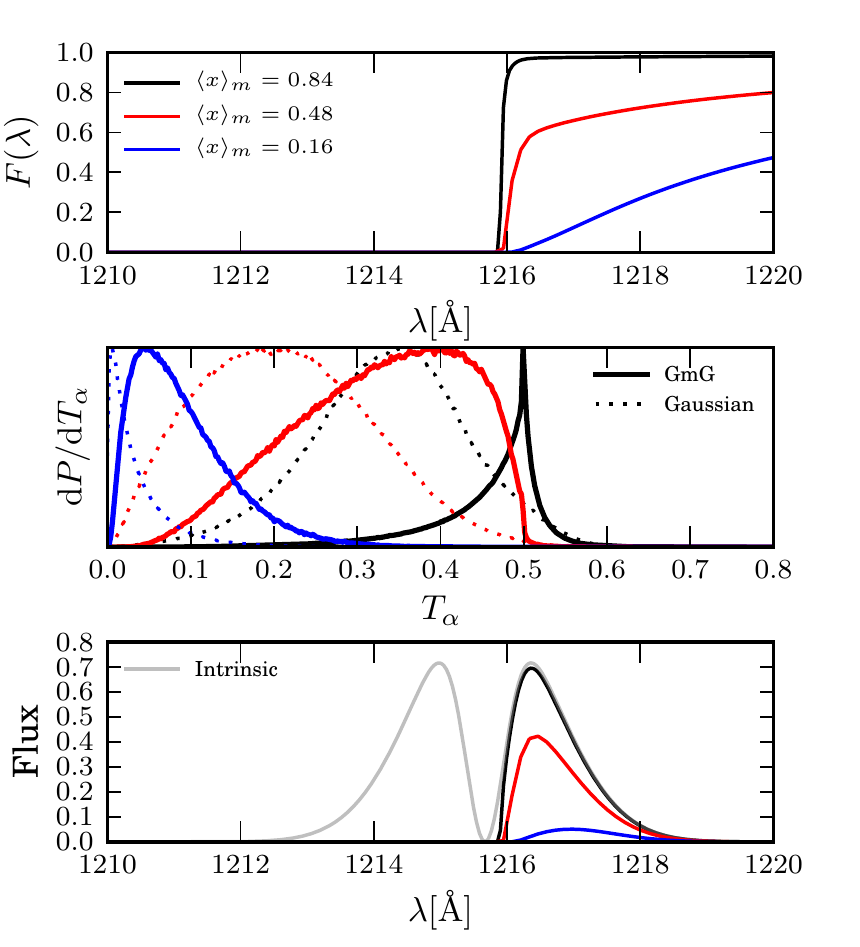}
	\caption{\emph{Top}: Median IGM transmission functions for varying ionized fractions. \emph{Middle}: probability density functions of the transmitted fraction of \Lya photons for all our simulated sightlines, i.e.\ 50 sightlines each for all haloes with $\Mhalo > 10^{10} \Msol$. The solid lines are for the GmG line profile and the dotted lines are for the Gaussian profile. The colour coding is the same as for the top figure. \emph{Bottom:} Examples of observed line shapes for a $10^{10} \Msol$ halo. The grey line shows the intrinsic line.}
	\label{fig:transmissions}
\end{figure}
\label{sec:general_properties}
Even a very small amount of neutral hydrogen is enough to absorb virtually all radiation bluewards of \Lya. For a highly ionized IGM, with only some very small amount of residual \hi, the transmission $F(\lambda)$ resembles a step function. At higher neutral fractions (and for certain individual sightlines in the more ionized cases) the transmissions begin showing damping wings that absorb some flux on the red side of the line centre, as illustrated in the top panel of Fig.\ \ref{fig:transmissions}. 

The middle panel of Fig.\ \ref{fig:transmissions} shows the evolution of the probability density function (PDF) of the transmitted fraction, $T_{\alpha}$, defined as:
\begin{equation}
	T_{\alpha} = \frac{\int J_{\mathrm{em}}(\lambda) F(\lambda) \ud \lambda}{\int J_{\mathrm{em}}(\lambda) \ud \lambda}.
	\label{eq:talpha}
\end{equation}

The PDFs were calculated over all the simulated sightlines, using the GmG line shape model (for comparison, we also show the results from the Gaussian line model). Note that $T_{\alpha}$ depends only on the \emph{shape} of the intrinsic line, and not on the value of the \Lya luminosity (we will discuss luminosities later on). Furthermore, since the blue peak is usually completely absorbed, only the red peak has an effect on the variation in $T_{\alpha}$ with damping wing strength. The blue peak only affects the absolute value of $T_{\alpha}$.

In a more neutral IGM, the strength of the damping wing is the most important factor for determining $T_{\alpha}$. For low ionized fractions, the damping wings are strong enough to absorb almost all the emitted photons, and the PDF peaks at only a few percent transmission. For intermediate $\xm$ the large variation in damping wing strength among different sightlines becomes apparent as the PDF becomes very broad. 

As we approach $\xm = 1$, the distribution of $T_{\alpha}$ clusters very strongly around 50\%. This effect is made stronger by the double-peaked nature of the line profile. At this point, any damping wings will be weak, and mostly affect the transmission close to the line centre. But since most of the emission is in the wings, a couple of \AA~ away from the centre, the transmitted line will consist of most of the red wing and nothing more.

\begin{figure}
	\begin{center}
		\includegraphics{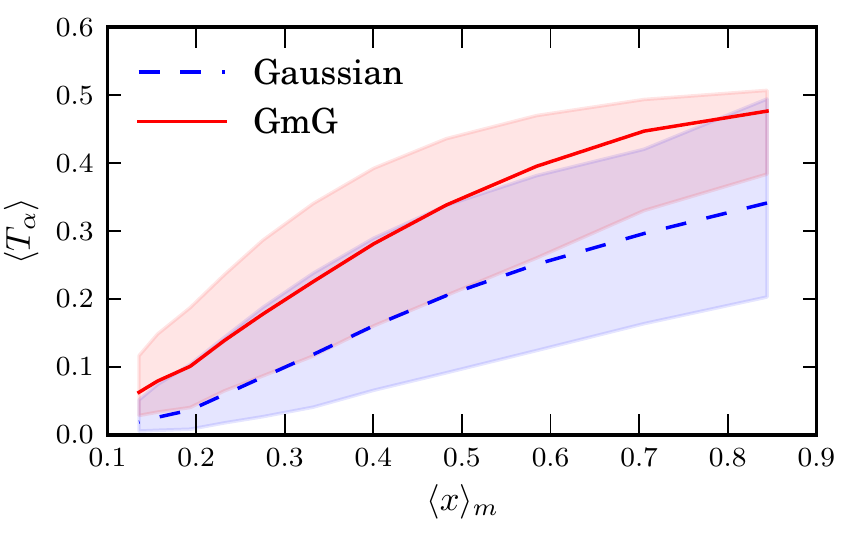}
	\end{center}
	\caption{Median $T_{\alpha}$ taken over all sightlines for all haloes as a function of IGM mean ionized fraction, for the Gaussian and GmG line models. The shaded areas show the $1\;\sigma$ spread among individual sightlines.}
	\label{fig:t_mean_hist}
\end{figure}

A single-peaked line profile centred on the \Lya line centre is more sensitive to damping wing strength than a line profile with most of the emission offset from the line centre, such as our GmG model. This results in wider $T_{\alpha}$ PDFs, as seen in the middle panel of Fig.\ \ref{fig:transmissions}. It also affects the average value of $T_{\alpha}$ as a function of $\xm$, as we show in Fig.\ \ref{fig:t_mean_hist}. The GmG model has a somewhat flatter relation between $\langle T_{\alpha} \rangle$ and $\xm$, especially at the later stages of the EoR. In other words, using a simplified model such as the single-peaked Gaussian may result in an over-prediction of the IGM absorption.

\subsubsection{$T_{\alpha}$ dependence on halo mass}
\label{sec:t_mass}

\begin{figure*}
	\centering
	\includegraphics[width=15cm]{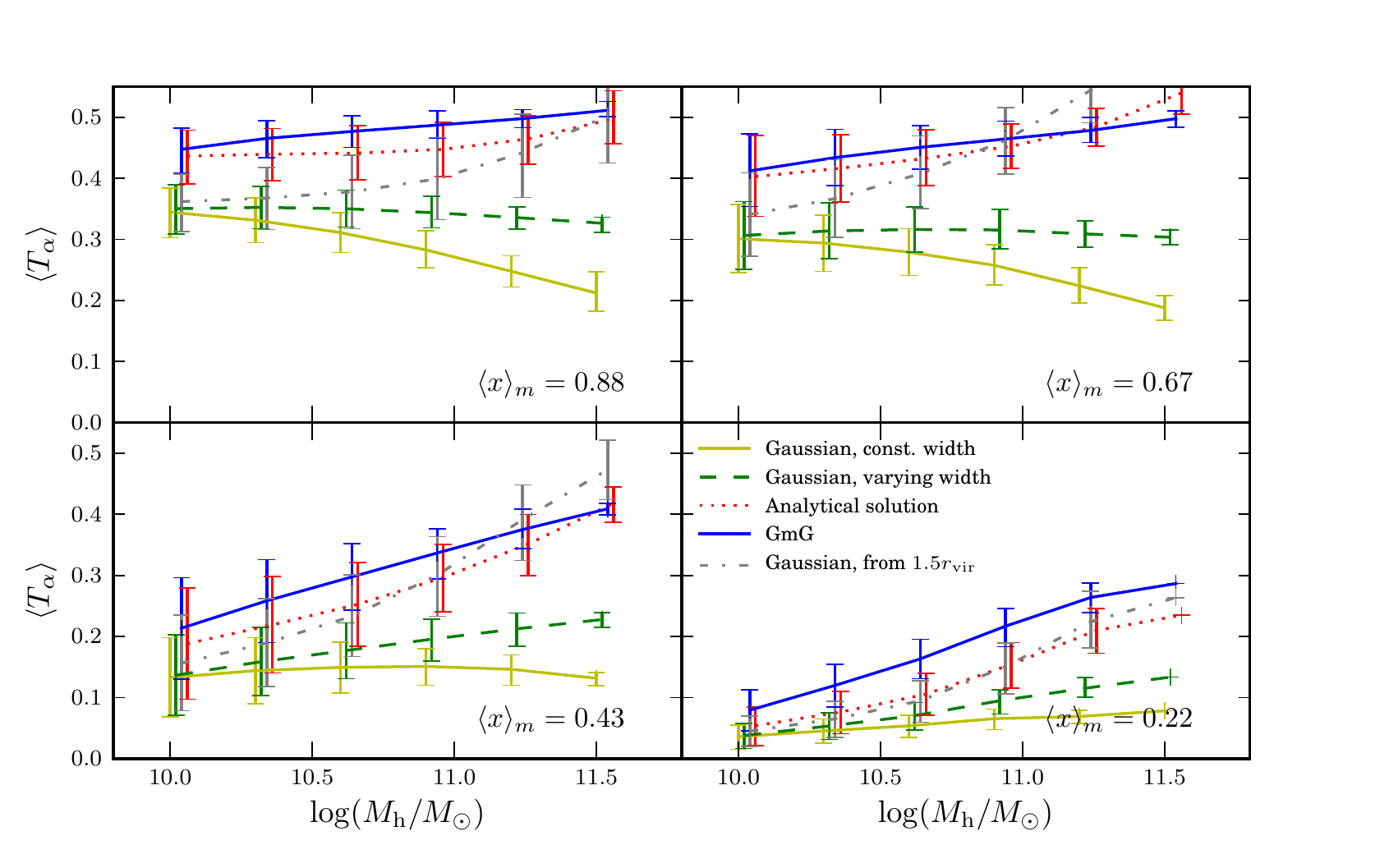}
	\caption{Mean IGM transmission for the various line models discussed in the text, binned according to halo mass, for four different IGM ionized fractions. The error bars show the $1\;\sigma$ spread of the $\langle T_{\alpha} \rangle$ distributions for each mass bin.}
	\label{fig:mass_trans}
\end{figure*}

While the global mean value of $T_{\alpha}$ at a given redshift is the most important factor in determining the change in LAE luminosity function and the decrease in the LAE fraction, the relationship between halo mass and $T_{\alpha}$ can affect a number of second-order effects, such as the shape of the LF and the IGM induced boost of the correlation function. This relationship is complex and depends on a number of factors. 

The strength of the damping wing depends on the density of neutral hydrogen in the close vicinity of the source, and so the transmitted fraction of \Lya will be a tracer of the size of the \hii region around the source. Since massive haloes tend to produce more ionizing photons, one would expect their surrounding \hii regions to be larger, leading to higher values for $T_{\alpha}$. Furthermore, massive haloes predominantly form in highly biased regions with many smaller-mass neighbours. These all contribute to enlarge the \hii region and, again, lead to larger $T_{\alpha}$. On the other hand, more massive haloes will tend to have more infalling gas, which tends to quench the blue portion of the \Lya spectrum and give a lower IGM transmission. Also, the biased regions with the most massive haloes will typically have a higher gas density, which --- if not highly ionized --- will lower the transmission.

Earlier studies have approached this problem in different ways, and come to different conclusions about the halo mass -- $T_{\alpha}$ relation. \cite{iliev2008a} use a simple Gaussian model and find that infalling gas gives a lower transmitted fraction for more massive haloes. \cite{mcquinn2007} perform similar simulations, but find almost no trend at all in two of their models, and an opposite trend in their model with the largest \hii regions. \cite{zheng2010} perform full radiative transfer calculations, and find a weak trend with lower $T_{\alpha}$ for higher-mass haloes, which they attribute to the higher density in the regions around these objects.

In Fig.~\ref{fig:mass_trans} we show the relation between $\langle T_{\alpha} \rangle$ and halo mass for four different line models. The two single Gaussian models refer to the models used in \cite{iliev2008a} --- one with a constant line width of $150$ km/s and one with a width of $150 \; \mathrm{km/s} \; \times [\Mhalo/(10^{10}\Msol)]^{1/3}$. The analytical solution refers to the line shape given in Eq.\ \eqref{analytical_line}. In addition, we show the results of using the single-width Gaussian line, but starting the sightlines $1.5$ virial radii outside the galaxies.

For our GmG model, there is a clear trend of higher $\langle T_{\alpha} \rangle$ for higher halo mass. This seems to contradict the results of \cite{iliev2008a}, who found a \emph{negative} correlation. This discrepancy is most likely due to the fact that they did not divide the \Lya radiative transfer into two regimes, as we discussed in Sec.\ \ref{sec:igmtransfer}. Since the amount of infall is largest close to the galaxies, this can not be studied with the $\exp(-\tau)$ model, but needs full radiative transfer. In our model, where we start the IGM radiative transfer at $1.5\; r_{\mathrm{vir}}$, we implicitly compensate for most of the infall effects when we fit the mass-to-light ratio in Sec.\ \ref{sec:intrinsic_luminosity}. 

The green and yellow lines of Fig.\ \ref{fig:mass_trans} show the results from using the Gaussian model, starting sightlines at the halo centres. Here, we do indeed see the same trends as \cite{iliev2008a}, with lower transmission for high-mass haloes at high $\xm$, especially for the single-width line model. However, if we start the Gaussian lines at the edge of the galaxies (grey dash-dotted lines), the infall trends disappear completely, indicating that the regions very close to the galaxies have a large effect on the transmission. Since these regions are not properly resolved in our large-scale simulation, one should be careful to draw any firm conclusions on the effects of infall based on these results. In our high-resolution galaxy sample, we saw no strong effects from infall, but higher resolution and better knowledge of input parameters are likely needed to properly study this. Also, different reionization models can give different results, as seen in \cite{mcquinn2007}. The reionization scenario used in \cite{iliev2008a} did not include self-regulation effects, and so one should be careful when comparing their results to the ones presented here.

\begin{figure}
	\centering
	\includegraphics[width=\columnwidth]{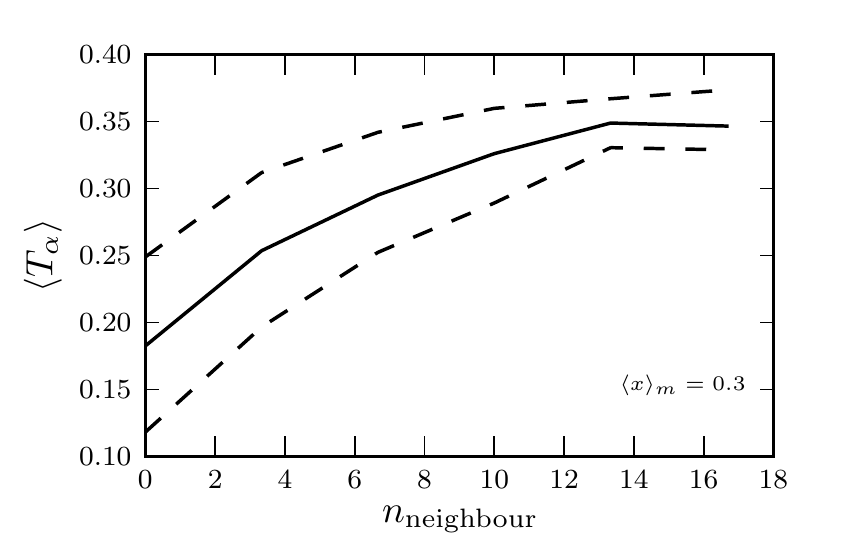}
	\caption{Correlation between mean $T_{\alpha}$ for a given halo and the number of close neighbour haloes ($d<5$ cMpc). Only haloes with $10 < \log \Mhalo/\Msol < 10.1$ were selected. The solid line shows the mean and the dotted lines show the $1\; \sigma$ spread for different bins.}
	\label{fig:neighbors_trans_corr}
\end{figure}

The main reason for the positive correlation seen when splitting the radiative transfer into two parts is that higher-mass haloes are usually located in larger \hii regions. This is due partly to the fact that large galaxies emit more ionizing radiation than smaller ones, and thus ionize the gas around them to a larger extent. However, it is likely more dependent on the fact that high-mass haloes tend to be located in biased regions with many neighbouring haloes surrounding them. These neighbours all help create large \hii regions, as was also discussed in \cite{iliev2008a}. In Fig.\ \ref{fig:neighbors_trans_corr} we show the correlation between $\langle T_{\alpha} \rangle$ and the number of close neighbour haloes (here defined as haloes within 5 cMpc, a distance which typically puts neighbours within the same \hii region) for haloes with $\log \Mhalo/\Msol \approx 10$. Since all these haloes have the same mass, they all ionize the IGM to the same extent, and have similar amounts of infall. Still there is a fairly strong correlation between number of neighbours and $\langle T_{\alpha} \rangle$.

The second thing to note in Fig.\ \ref{fig:mass_trans} is that even if we fix the starting point of the sightlines to $1.5\;r_{\mathrm{vir}}$, the choice of line model has an appreciable effect on the results. As we discussed above, lines with a lot of emission at the line centre are more sensitive to damping wing strength, while the double-peaked models (GmG, analytical) tend to give somewhat flatter relations between $\langle T_{\alpha}$ and halo mass.

In summary, the exact nature of the halo mass-$T_{\alpha}$ relation is complicated, and will depend on the definition of $T_{\alpha}$ and assumptions about the \Lya line shape. Here, we are interested in the effects of the IGM, and define $T_{\alpha}$ as the transmission through the IGM \emph{after the photons leave the galaxies}. For this definition, we see a fairly weak trend with higher transmission for higher-mass haloes. In our model, any infall effects that are taking place close to the galaxies will be calibrated away when we fit the luminosities later on.


\subsection{LAE luminosity functions}
Observing the evolution of the LAE luminosity function (LF) with redshift can provide information about both the change in intrinsic properties of the galaxies and the changing state of the surrounding IGM. Unfortunately, it is far from straightforward to disentangle these effects. For example, a decrease in the LF could come from a decrease in star formation, a higher dust content, a denser and/or more neutral IGM or a combination of all three.

The LAE LF appears to change very little, if at all, up to around $z=6$ \citep{malhotrarhoads2004,ouchi2008,tilvi2010}. In other words, if the reionization of the Universe is still ongoing at these redshifts, the decrease in observability of LAEs due to the neutral IGM must be counteracted by an intrinsic change of the galaxy properties, such as a lower dust content.

However, at higher redshifts there is an observed change in the LF. \cite{ouchi2010} (hereafter Ou10) observe a sample of 207 LAEs at $z=6.5$ in the Subaru/XMM-Newton Deep Survey (SXDS) field, and find that the LF at $z=6.5$ is different from the observations at $z=5.7$ with a fairly high degree of certainty. By comparing to simulations by \cite{mcquinn2007}, \cite{iliev2008a}, \cite{dijkstra2007} and \cite{furlanetto2006} they conclude that this suggests a change in neutral fraction of the IGM of $\Delta \xm \approx 0.2$. This value assumes a model for the intrinsic evolution of LAEs based on observed changes in the UV LF.

\cite{kashikawa2006} and \cite{kashikawa2011} (hereafter Ka11) also observe the LAE LF at $z=6.5$ in the Subaru Deep Field (which is separate from the SXDS field of Ou10) and see a difference compared to $z=5.7$. Comparing their results to models both by \cite{mcquinn2007} --- who do not include any intrinsic galaxy evolution --- and \cite{kobayashi2010} --- who do model intrinsic evolution --- they require $\Delta \xm \sim 0.4$ at $z=6.5$. This assumption is motivated by the fact that they, in contrast to Ou10, do not see any evolution in the UV LF during the same time interval. \cite{dayal2008}, are able to reproduce the LF evolution between $z=6$ and $z=6.5$ with only galaxy mass evolution, but this then requires a very sharp drop in \Lya escape fraction to explain the absence of change in the LF at $z<6$. 

\subsubsection{Intrinsic luminosity model}
\label{sec:intrinsic_luminosity}

We now investigate for which ionized fraction our models best reproduce the observed LFs by Ou10 and Ka11. For this we need a model for the \Lya luminosity of our simulated haloes. \cite{iliev2008a} and \cite{mcquinn2007} assume a constant ratio between dark matter halo mass and \Lya luminosity, and fit this to the observations. In reality, however, we expect there to be some variation in luminosity among haloes of the same mass, which is also what we see in the detailed simulations of our high-resolution galaxy sample. Not only is there not a direct one-to-one relation between halo mass and production of \Lya photons, but the escape fraction can vary a lot between different sightlines \citep{laursen2007}. While a constant mass-to-light ratio will give a LAE LF with a shape very similar to the mass function (the difference coming only from the dependence of $\langle T_{\alpha} \rangle$ on halo mass), adding some random scatter to this relation will tend to flatten the LF and give a higher bright end and a lower faint end. 

\begin{figure}
	\begin{center}
		\includegraphics{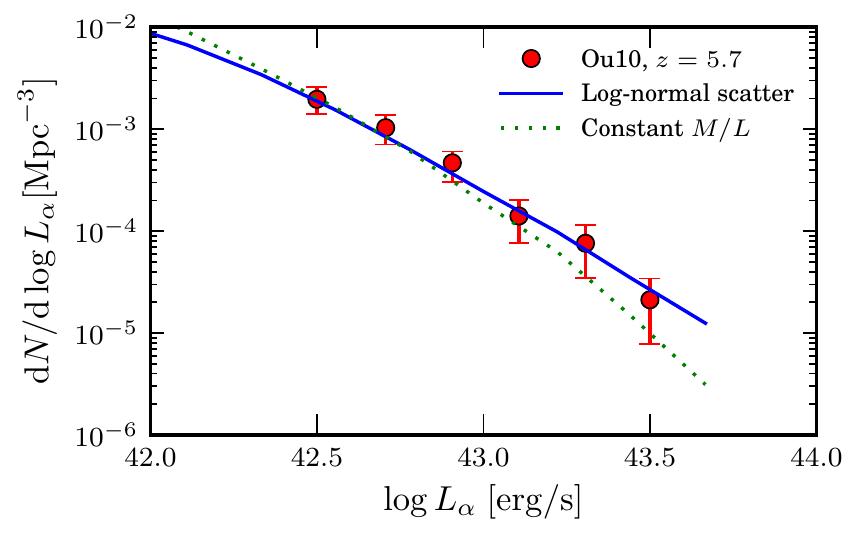}
	\end{center}
	\caption{Simulated \Lya LFs compared to observations at $z=5.7$ by Ou10. The dotted green line shows the best fit assuming a constant mass-to-light ratio. The blue line is the best fit assuming a log-normal scatter in the mass-to-light.}
	\label{fig:lae_lf_5p7}
\end{figure}

We attempt to find a good model for the \Lya luminosity by fitting our results to the observed LF at $z=5.7$ from Ou10. Our IGM simulation does not go down to such low redshifts, so we take our lowest-redshift halo sample, from $z=6.0$, and make the assumption that $\langle T_{\alpha}\rangle=0.5$ for all haloes at this redshift. This is motivated by Fig.\ \ref{fig:transmissions}, where it is shown that the $T_{\alpha}$ distribution becomes very narrow and centred on $0.5$ for low neutral fractions (as is shown in \cite{songaila2004}, there is still enough residual \hi to remove the blue part of the spectrum completely). 

In Fig.\ \ref{fig:lae_lf_5p7} we show our best fit to the observations, assuming a model where the \Lya luminosities of haloes of a given mass follow a log-normal probability distribution with a mean that is proportional to the halo mass, and a fixed width. When fitting this relation to the observations, we fixed the scatter to $\sigma=0.4$ (see Eq.\ \ref{mass_to_light}). This value and the log-normal distribution were motivated by the simulations of the high-resolution galaxy sample discussed in Sec.\ \ref{sec:linemodel}, and are similar to the model used by \cite{munoz2011}. The constant mass-to-light model is roughly consistent with the data, but appears to give slightly too steep a bright end. The log-normal-scatter model, however, matches the observations very well. By $\chi^2$ minimisation on a grid, we find that the best fit to the observations is:
\begin{align}
	\nonumber
	& \log L_{\alpha} \; [\mathrm{erg/s}] = \\
	&= \log (\Mhalo/\Msol) + \mathrm{norm(\mu = 31.7, \sigma = 0.4)},
	\label{mass_to_light}
\end{align}
where $\mathrm{norm}(\sigma,\mu)$ denotes a random variable distributed according to a normal distribution with mean value $\mu$ and width $\sigma$.

We would like to point out that the value of $\mu$ in Eq.\ \eqref{mass_to_light} should not be interpreted too strictly. As we discussed earlier, there is a degeneracy between the absolute value of $T_{\alpha}$ and the intrinsic luminosity. Had we used a line model with a stronger red peak, our $T_{\alpha}$ values would be higher, and, consequently, the intrinsic luminosities lower. In this study, we are only concerned with the dependence of $T_{\alpha}$ on, for instance, IGM neutral fraction and halo mass, and so this degeneracy has little effect on our results.

\subsubsection{Comparison to observations at $z=6.5$}
\label{sec:lumfunc_comparison}
\begin{figure*}
	\begin{center}
		\includegraphics[width=13cm]{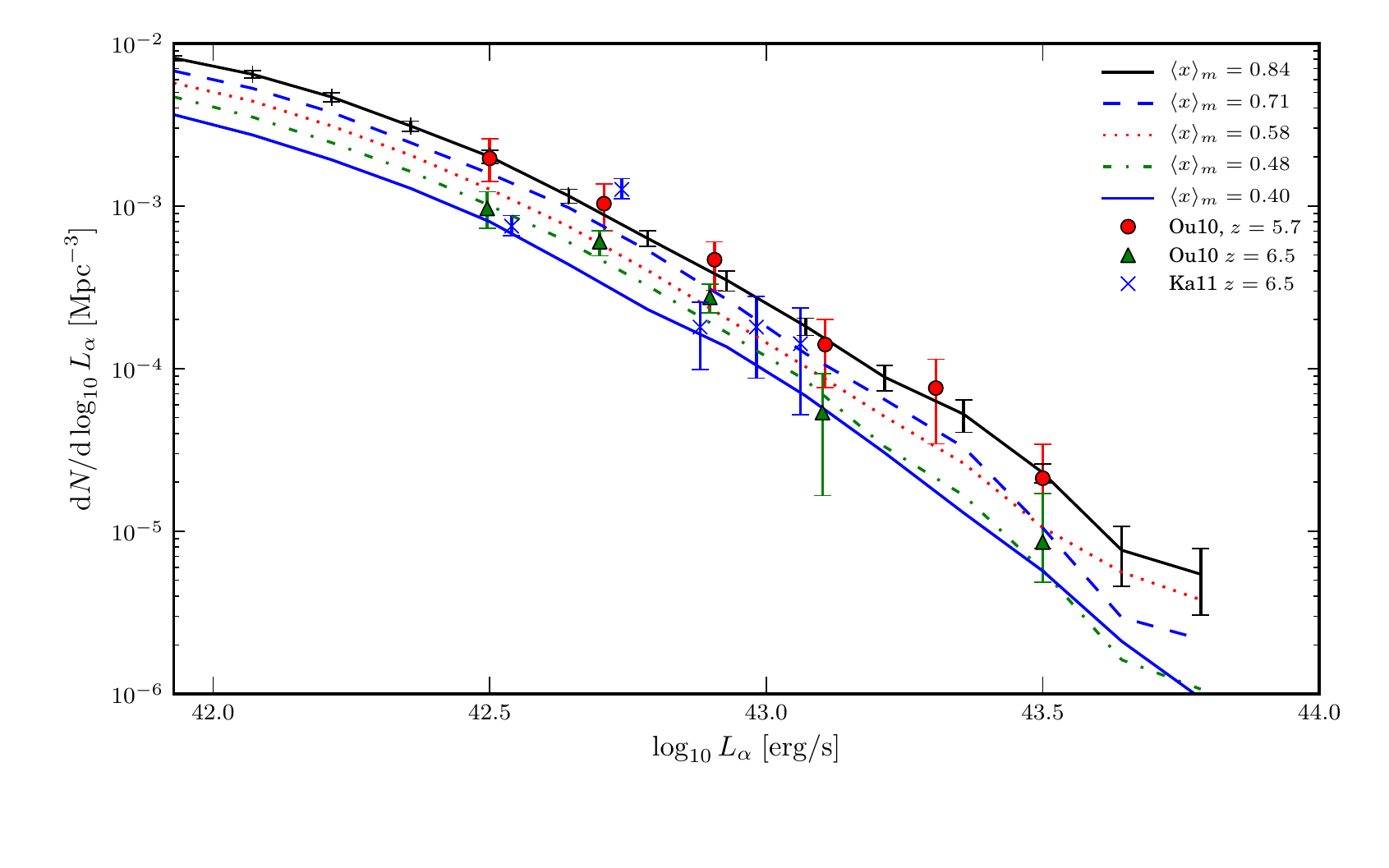}
	\end{center}
	\caption{Simulated \Lya LFs for five different IGM ionization fractions, compared to observations at $z=5.7$ and $z=6.5$. Poisson errors are shown for the first LF; they have been omitted from the others for clarity. All LFs were calculated assuming no intrinsic evolution in the LAE luminosities.}
	\label{fig:lae_lfs_6p5}
\end{figure*}

In Fig.\ \ref{fig:lae_lfs_6p5} we show simulated LFs for different IGM ionization states, compared to observations at $z\sim6.5$ by Ou10 and Ka11. It is evident from this figure that the uncertainties in the observations are too large to put any robust constraints on $\xm$ at $z=6.5$. Nevertheless, we still investigate for which $\xm$ our simulations best fit the data.

To do this, we make the simplifying assumption that the underlying intrinsic LAE LF does not change between $z=6$ and $z=6.5$, in accordance with the observations at lower redshifts discussed above. For this purpose, in producing Fig.\ \ref{fig:lae_lfs_6p5}, we have scaled the underlying halo masses in such a way as to give a constant LAE mass function at all ionization states. This ensures that all the evolution in the LF comes from the changing ionization state of the IGM. We then go on to calculate the $\chi^2$ goodness-of-fit, comparing the LFs in Fig.\ \ref{fig:lae_lfs_6p5} to the observations by Ou10 and Ka11. The best fit is for an IGM ionized fraction of around $\xm \sim 0.5$, but the errors in the data are still too large to get a robust constraint. However, it is interesting to note that both the Ou10 and Ka11 samples, which come from different fields, both give the same best fit value.

It is worth pointing out here, that while we look for the best-fitting ionized fraction at $z=6.5$, this value for $\xm$ occurs already at $z=7.2$ in our simulations (see Fig.\ \ref{fig:ionization_hist}). This makes the comparison somewhat inconsistent, and implies that reionization in fact took place later than in our model. However, we believe the comparison is still reasonable, since shifting our IGM simulations in time would not significantly alter the results, as we argued above.

The value of $\xm \sim 0.5$ at $z=6.5$ is in conflict with other measurements as we will discuss in Sec.\ \ref{sec:discussion}, which casts some doubt on the assumption that the drop in LF is due only to an increasingly neutral IGM. If we drop this assumption and allow the mass of the DM haloes to evolve according to our N-body simulations while assuming that Eq.\ \eqref{mass_to_light} holds at all times (i.e.\ there is no change in dust content, star formation efficiency etc.), then we find that the drop in LF can be explained with an IGM that is completely reionized already at $z>6.5$. This, however, would imply that the LF continues to grow in amplitude at $z<6$, which is incompatible with observations. It would seem most likely that the LF drop is due to a combination of IGM and intrinsic evolution, but observations of LAEs alone can not break this degeneracy.

\subsubsection{Sample variance}
\label{sec:sample_variance}

In the figures below, we illustrate the effect of cosmic variance and statistical fluctuations when observing in a limited field. We do this by considering three different observational boxes:
\begin{itemize}
	\item The full box. This is our entire simulated $163$ cMpc box, where we assume we have the full 3D positions of all the haloes. 

	\item A deep box. An optimistic approximation of what might be observed in a future space-based survey, e.g.\ with the JWST. Such a survey will have trouble achieving a large field-of-view, but is unconstrained by atmospheric effects and so can, given enough time, obtain spectroscopy over a large redshift range. We assume this box is $60$ arcmin$^2$, $\sim1/6$ the side of the full box, with the depth set to the whole box depth. We also assume, optimistically, that we have spectroscopic data, i.e.\ 3D positions, of all the galaxies in the box.

	\item A thin box. This approximates a large ground-based photometric survey with a large field-of-view but limited redshift range. We assume a field of $1$deg$^2$ which is approximately the size of our full box, with $\Delta z = 0.1$, i.e.\ $1/5$ the depth of the full box. Here we assume that we only know the 2D positions of the haloes. This is roughly equivalent to the Subaru Deep Field (Ka11).
\end{itemize}

\begin{figure*}
	\centering
	\includegraphics{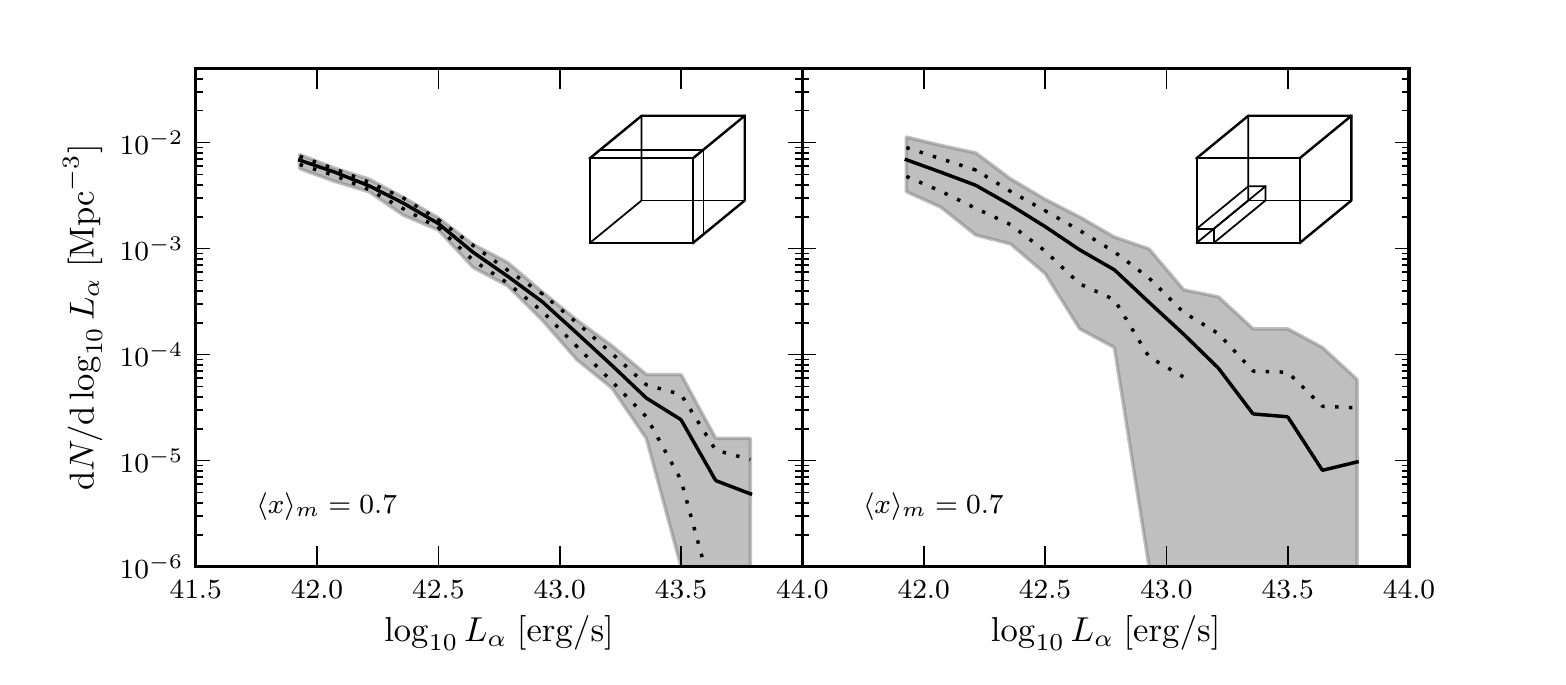}
	\caption{Left panel: Variation in LAE LFs from several thin box subfields taken from our big simulation box. The solid black line shows the mean value for the LF, the dotted lines show the $1\sigma$ variation and the shaded area indicates the range between the minimum and maximum values of the LF for each luminosity bin. Right panel: Same as left, but for the deep sub box.}
	\label{fig:subbox_lfs}
\end{figure*}

\begin{figure*}
	\begin{center}
		\includegraphics[width=17cm]{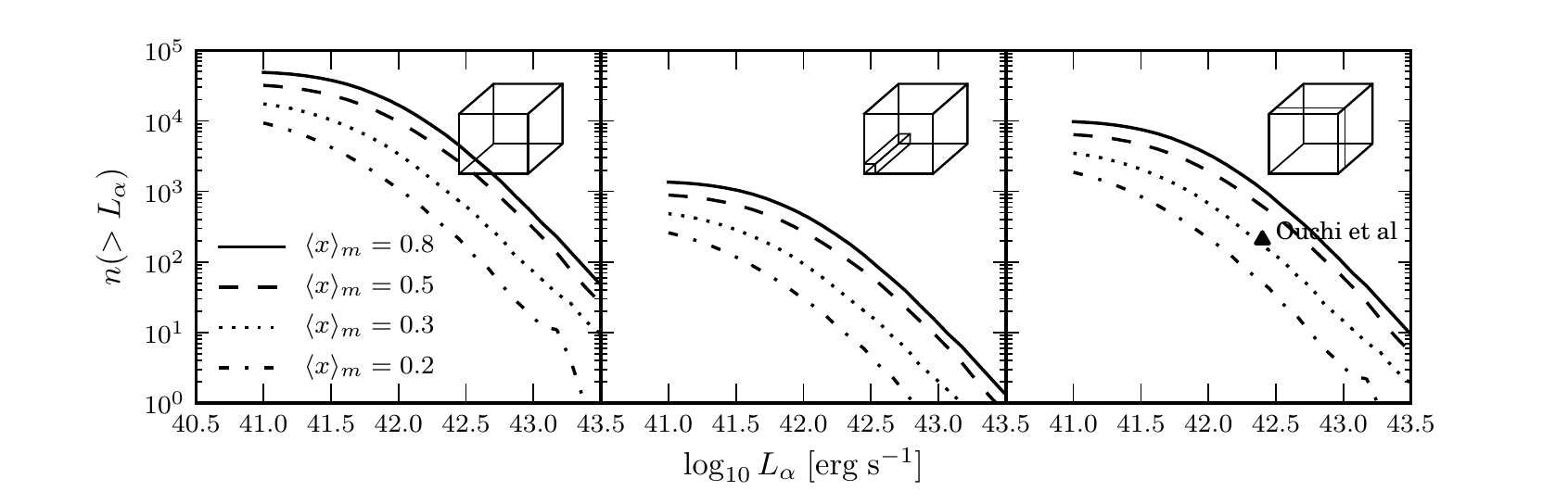}
	\end{center}
	\caption{Number of LAEs above a given luminosity limit in our full box and the two subboxes for different ionized fractions. As a reference point, we show in the rightmost panel the sample of \protect\cite{ouchi2010}, who detect 207 LAEs with $L_{\alpha} > 2.5\times10^{42}$ erg/s at $z=6.5$, in a field roughly the same size as our thin box.}
	\label{fig:detectability}
\end{figure*}

Fig.\ \ref{fig:subbox_lfs} shows the field-to-field variation in LAE LFs. The figure was produced by splitting our simulation box into many small sub-boxes, and calculating the LAE LF for each sub-box. It is clear that sample variance becomes a major issue in a small field such as our deep box. Here, the field-to-field variation is very large compared to the change in LF due to an increasing neutral fraction, as seen in Fig.\ \ref{fig:lae_lfs_6p5}, and in many cases the higher-luminosity bins end up empty, making the bright end very difficult to study. However, for the thin box, the field-to-field variation is of the same order as a $\sim 10$\% change in $\xm$, at least at the faint and intermediate parts of the LF. In Fig.\ \ref{fig:detectability} we show the number of LAEs above a given luminosity in the full box and in each of the sub-boxes.

\subsection{LAE fraction}
Several observers report a sharp decrease in the fraction \xa of drop-out selected galaxies that show \Lya emission between redshifts 6 and 7 \citep{ota2008,stark2010,pentericci2011,ono2011,schenker2012}. For example, \cite{pentericci2011} observe a sample of 20 drop-out selected galaxies, and find \Lya emission in only 3, and \cite{schenker2012} find only 2 LAEs (plus one possible) in a sample of 19 drop-outs. While there are a number of effects that could explain this, arguably the most obvious one would be an increase in the IGM neutral fraction, as noted by e.g.\ \cite{romero2012}. 

\begin{figure}
	\begin{center}
		\includegraphics{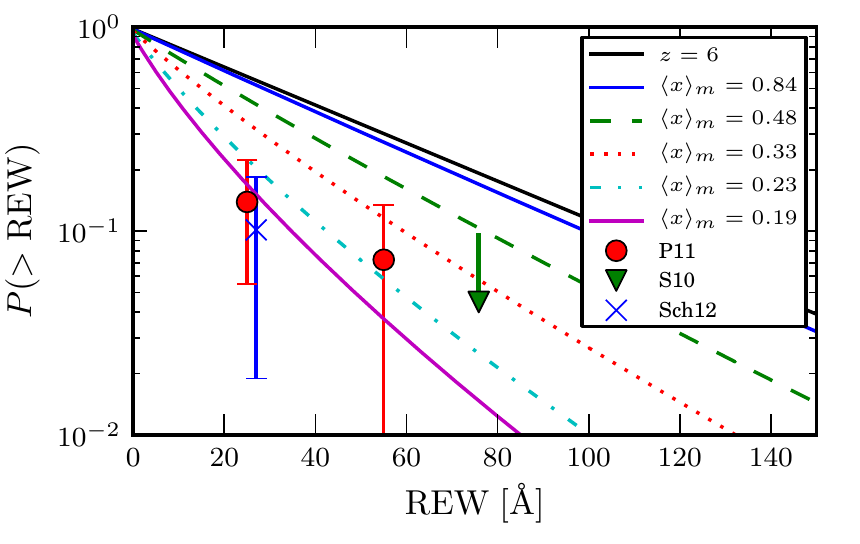}
	\end{center}
	\caption{Fraction of LAEs above a given REW for different IGM ionization states, compared to observations at $z=7$. The solid black line is for $z=6$, assuming $\langle T_{\alpha} \rangle = 0.5$. The observations are: P11 = \protect\cite{pentericci2011}; S10 = \protect\cite{stark2010}; Sch12 = \protect\cite{schenker2012}.}
	\label{fig:rew_cdf_t0p5}
\end{figure}

\begin{figure}
	\begin{center}
		\includegraphics{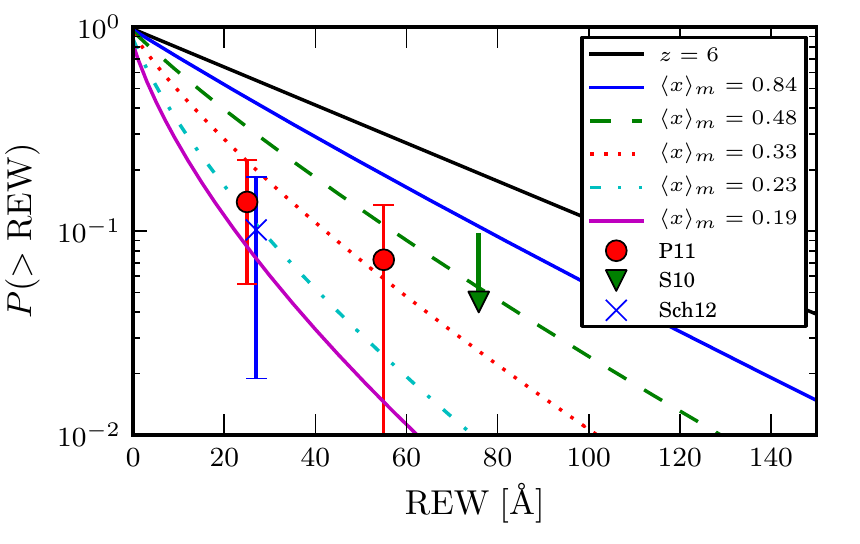}
	\end{center}
	\caption{Same as Fig.\ \protect\ref{fig:rew_cdf_t0p5}, but using the simple Gaussian model.}
	\label{fig:rew_cdf_gauss}
\end{figure}

Here, we follow the procedure of \cite{dijkstra2011} to estimate the decrease in \xa for an increasing neutral fraction, using the distribution of observed \Lya rest frame equivalent widths (REW). We begin by assuming that the REWs at $z=6$ are distributed according to an exponential distribution $P_{z=6}(\mathrm{REW}) \propto \exp\left( \frac{-\mathrm{REW}}{\mathrm{REW}_c} \right)$ and setting $\mathrm{REW}_c = 50$ \AA~to match the observations at this redshift \citep{stark2010}. Making again the simplifying assumption of no intrinsic LAE evolution, the distribution at some other redshift $z$ can be written as:
\begin{equation}
	P_z(\mathrm{REW}) = N \int_0^1 P_{\mathrm{intr}}(\mathrm{REW}/T_{\alpha}) P_z(T_{\alpha})\ud T_{\alpha},
	\label{prew}
\end{equation}
where $P_z(T_{\alpha})$ is the PDF of the \Lya transmitted fraction, shown in Fig.\ \ref{fig:transmissions}, $N$ is a normalisation constant and $P_{\mathrm{intr}}$ is the $\mathrm{REW}$ distribution for $T_{\alpha}$ = 1. 

\cite{dijkstra2011} assume, conservatively, that the IGM transmission is 100\% at $z=6$, i.e.\ that $P_{z=6}(\mathrm{REW}) = P_{\mathrm{intr}}(\mathrm{REW})$. Since observations show close to zero transmission on the blue side \citep{songaila2004}, this assumption is only valid if the intrinsic lines are very heavily biased towards the red wing, which is true in the model of \cite{dijkstra2011}, but not for our GmG model.  

To keep consistency with our previous discussion in Sec.\ \ref{sec:intrinsic_luminosity}, we assume that the IGM is very close to fully ionized at $z=6$, but still retains a small amount of residual \hi that absorbs the blue wing of the \Lya line, so that the $T_{\alpha}$ PDF takes the form $P_{z=6}(T_{\alpha}) = \delta(T_{\alpha}-0.5)$. This assumption is actually very similar to that of \cite{dijkstra2011}, since their emission is mostly at the red side of the line centre. Inserting this into Eq.\ \eqref{prew} gives the intrinsic equivalent width distribution:
\begin{align}
	\nonumber
	&P_{z=6}(\mathrm{REW}) = \\ \nonumber
	&= N \int_0^1 P_{\mathrm{intr}}(\mathrm{REW}/T_{\alpha})\delta(T_{\alpha}-0.5)\ud T_{\alpha} = \\
	&= N P_{\mathrm{intr}}(\mathrm{REW}/0.5),
	\label{intrprew}
\end{align}
so that:
\begin{equation}
P_{\mathrm{intr}}(\mathrm{REW}) \propto \exp\left( \frac{0.5\mathrm{REW}}{\mathrm{REW}_c} \right).
\end{equation}

Using this expression for $P_{\mathrm{intr}}(\mathrm{REW})$ we can calculate the REW distribution at higher redshifts using Eq.\ \eqref{prew}. Fig.\ \ref{fig:rew_cdf_t0p5} shows the expected number of LAEs above a given REW along with data points from a number of observations. \cite{pentericci2011} claim that they need a change in neutral fraction of $\Delta x_{\mathrm{HI}} \sim 0.6$ between $z=6$ and $z=7$ to explain the drop in \xa, and Fig.\ \ref{fig:rew_cdf_t0p5} seems consistent in requiring such an extremely fast ionization of the IGM. Even at $\xm$ as low as 0.3, the IGM absorption is just barely enough to be consistent with the observations. However, just as the luminosity functions, interpretation of the REW distribution is strongly dependent on the (possible) intrinsic galaxy evolution.

Another reason that we require such a low ionized fraction is our GmG line model. The fact that the \Lya photons mainly escape their host galaxies away from the line centre means that changes in the IGM damping wings have a smaller effect on $T_{\alpha}$ than for a line profile with emission at the line centre, as previously discussed. For reference, we show the REW cumulative distribution function (CDF) for the simple Gaussian model in Fig.\ \ref{fig:rew_cdf_gauss}. While not radically different from Fig.\ \ref{fig:rew_cdf_t0p5}, the Gaussian model does give a greater change in the CDF when adjusting the ionized fraction of the IGM.

%
\subsection{Correlation functions}
\begin{figure}
	\centering
	\includegraphics[width=\columnwidth]{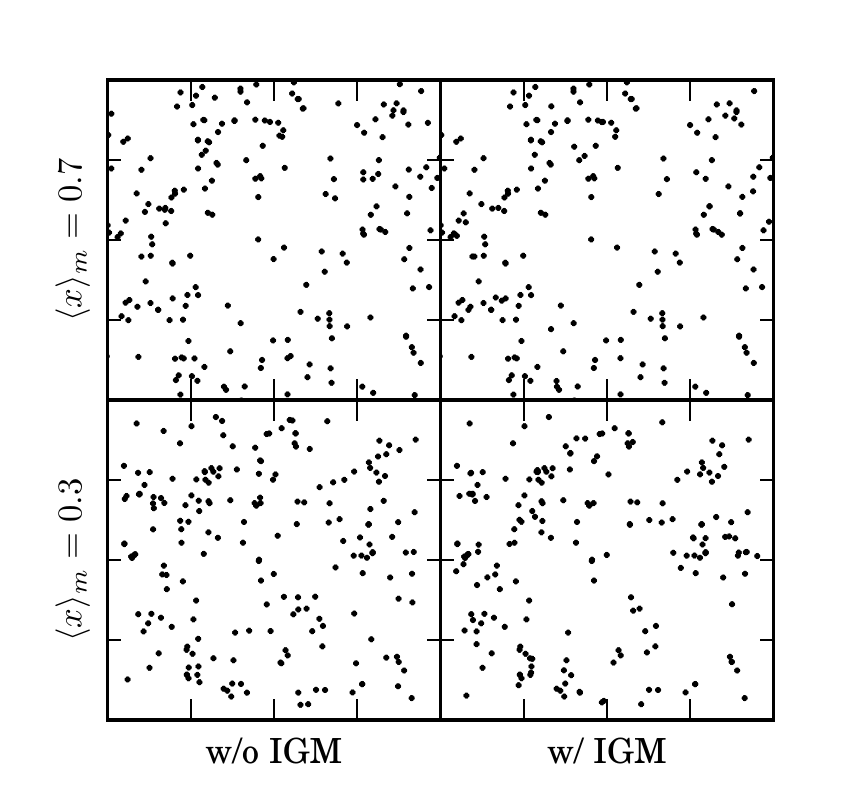}
	\caption{Projected positions for the 200 brightest haloes in our sample without (left) and with (right) IGM effects included. At a global ionization fraction of $\xm=0.7$, there is barely any noticeable increase in clustering. For $\xm=0.3$, however, the extra clustering starts to become visible.}
	\label{fig:halo_positions}
\end{figure}

The clustering of \Lya emitters has been proposed as a potential probe of the ionization state of the IGM (\citealt{mcquinn2007}). The reasoning is that on top of the intrinsic clustering of galaxies, the presence of large neutral patches of hydrogen which absorb the \Lya emission in some areas will cause the LAEs to appear more clustered. 

This is illustrated in Fig.\ \ref{fig:halo_positions}, which shows the positions of the most luminous haloes with and without the IGM radiative transfer applied. Since the ionized bubbles will be located around clusters of galaxies, the IGM acts as a kind of filter, obscuring haloes in non-biased regions, and showing only the most densely clustered regions. Thus, by comparing the clustering from a sample of LAEs to a sample of galaxies selected by some other method that is not sensitive to the IGM, it should be possible to say something about the ionized state of the IGM. This method has the advantage of not requiring any assumptions regarding the intrinsic evolution of LAEs, only that any such evolution is independent of environment.

The quantitative effects of the IGM on the correlation function was studied by \cite{mcquinn2007}, who found that the effect should be large enough to be measurable in future large LAE surveys, using a simple Gaussian line model. \cite{iliev2008a}, however, concluded that the difference between the IGM and non-IGM case is very small and difficult to detect. In the preparation of this paper, we discovered that due to a data handling error the correlation functions shown in Figs.~27 and 28 of \cite{iliev2008a} incorrectly showed a much smaller difference than was actually the case. When plotting the correct data we find that the results in \cite{iliev2008a} were qualitatively consistent with those of \cite{mcquinn2007} and those shown in this paper.

\cite{kashikawa2006} calculate the two-point angular correlation function of 58 LAEs observed at $z=6.5$ in the Subaru Deep Field, but find no evidence of even intrinsic galaxy clustering, although they note that the small sample size and the relatively small surveyed area introduce large uncertainties. Also \cite{mcquinn2007} study the same sample and find it to be consistent with a fully ionized IGM. \cite{ouchi2010} measure the correlation function of 207 LAEs at $z=6.5$, but while they do detect a clustering in their sample, the signal is not strong enough to be attributed to the IGM.

In this section we investigate the potential of future observations to constrain the ionized fraction of the IGM using clustering measurements. We use the spatial two-point correlation function (2PCF) $\xi(r)$, which is the excess probability of finding two galaxies in two volume elements $\ud V_1$ and $\ud V_2$ separated by a distance $r$:
\begin{equation}
	\ud P = n^2[1 + \xi(r)] \ud V_1 \ud V_2
	\label{eq:correlation}
\end{equation}
where $n$ is the number density of galaxies. We calculate $\xi(r)$ using the method described in \cite{martel1991}. The projected version of $\xi(r)$ --- the two-point angular correlation function, usually denoted $w(\theta)$ --- is defined as:
\begin{equation}
	\ud P = n^2[1 + w(\theta)] \ud \Omega_1 \ud \Omega_2
	\label{eq:angular_correlation}
\end{equation}
where $\ud \Omega_1$ and $\ud \Omega_2$ are two solid angle elements separated by an angle $\theta$ on the sky. This function can be calculated from $\xi(r)$ using the Limber relation \citep{limber1953}, or directly using, for instance, the Landy-Szalay estimator \citep{landyszalay93}, which is the approach we use here. The 3D 2PCF obviously contains more information than the 2D version, but it requires knowledge of the line-of-sight positions of the galaxies, which at these redshifts can only be obtained from spectroscopy. 

Let us assume that a future hypothetical survey observed $N$ LAEs in a region the same size as our simulation box. Would it be possible to say something about the ionized state of the IGM from the correlation function of the galaxies alone?  In other words, is the correlation function in our model with the IGM included significantly different from the same model without the IGM taken into account?

\begin{figure}
	\centering
	\includegraphics[width=\columnwidth]{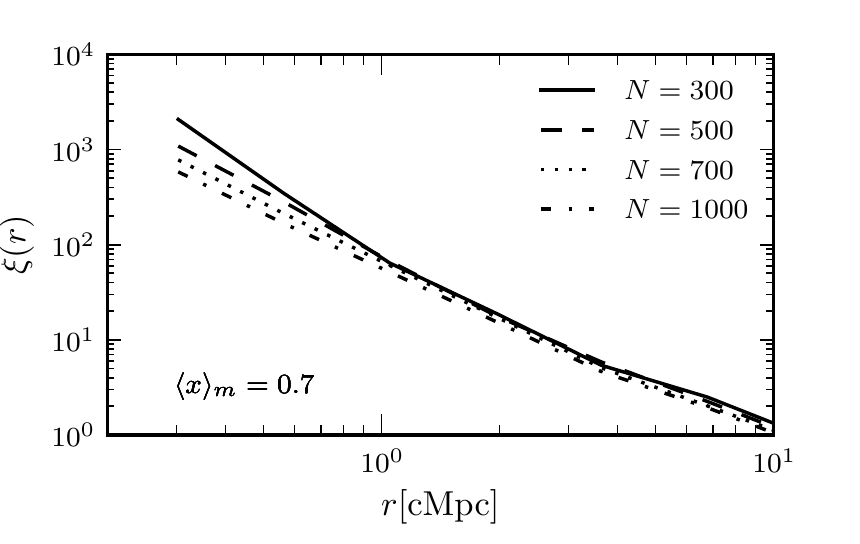}
	\caption{Mean 3D 2PCFs for the full simulation box with varying number of observed sources. Despite the IGM being held constant, there is a decrease in clustering when increasing the sample size, since more low-mass haloes from less biased regions are included.}
	\label{fig:correlations_vary_N}
\end{figure}

It might appear most intuitive at first to simply compare two samples (with and without IGM) with the same detection limit. This does indeed give a significantly higher correlation for the IGM case, but the difference in correlation comes only partly from the patchiness of the IGM, but mostly from the fact that the non-IGM sample will be much larger and include also some lower-mass haloes, which are less intrinsically clustered (illustrated in Fig.\ \protect\ref{fig:correlations_vary_N}).

In this study, we do not focus on the detection limits of our hypothetical surveys. Instead, we take the same approach as \cite{iliev2008a}, and simply assume that we observe a given number $N$ LAEs in each field. We then compare the 2PCF of this sample to that of a sample of equal size, but without the effects of the IGM taken into account (such a sample could be obtained by observing in a wavelength range that is not affected by the IGM). This way, we isolate the extra clustering that is due to the topology of the IGM, and not related to the decreased number density of sources.

\subsubsection{Simulation results}
\begin{figure}
	\centering
	\includegraphics[width=\columnwidth]{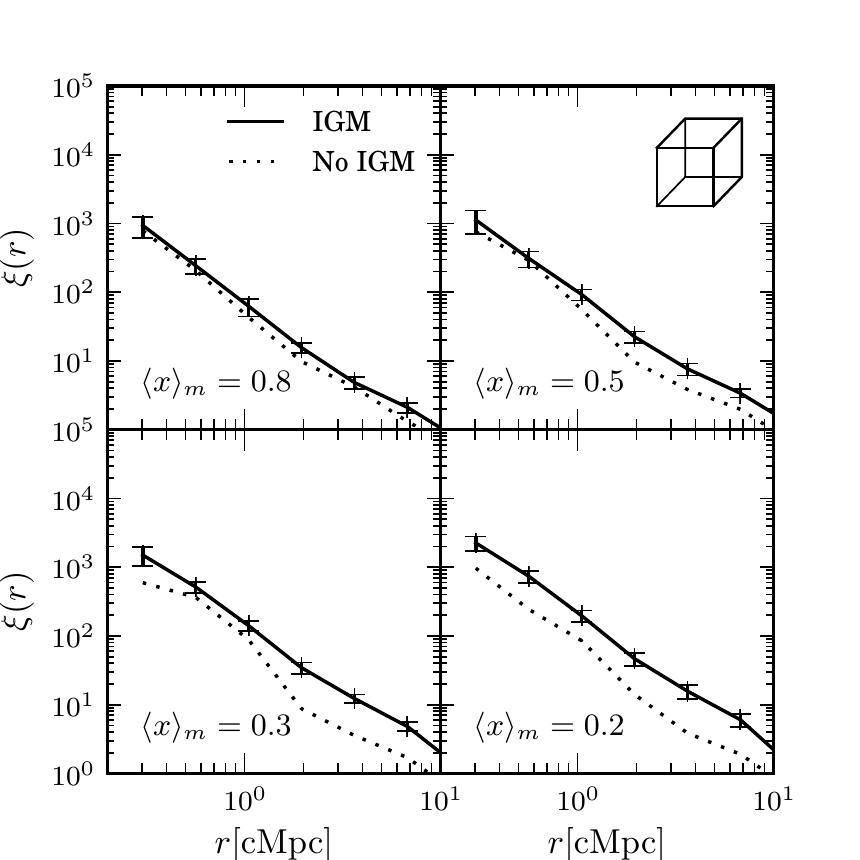}
	\caption{Two-point correlation functions for the 500 intrinsically brightest LAEs compared to the 500 brightest after including the IGM. The lines show the mean correlation function calculated from 50 random sightlines using the whole simulation box. The error bars indicate the $1 \sigma$ spread. }
	\label{fig:correlations_wholebox}
\end{figure}

Fig.\ \ref{fig:correlations_wholebox} shows the 3D 2PCFs calculated from the 500 intrinsically brightest LAEs (averaged over all 50 sightlines per source) in our full simulation box compared to the 500 brightest with IGM effects included, assuming the GmG line model and the log-normal luminosity relation, Eq.\ \eqref{mass_to_light}. The solid line shows the mean correlation function over 50 sightlines and the error bars show the $1 \sigma$ spread among sightlines.

\begin{figure}
	\centering
	\includegraphics[width=\columnwidth]{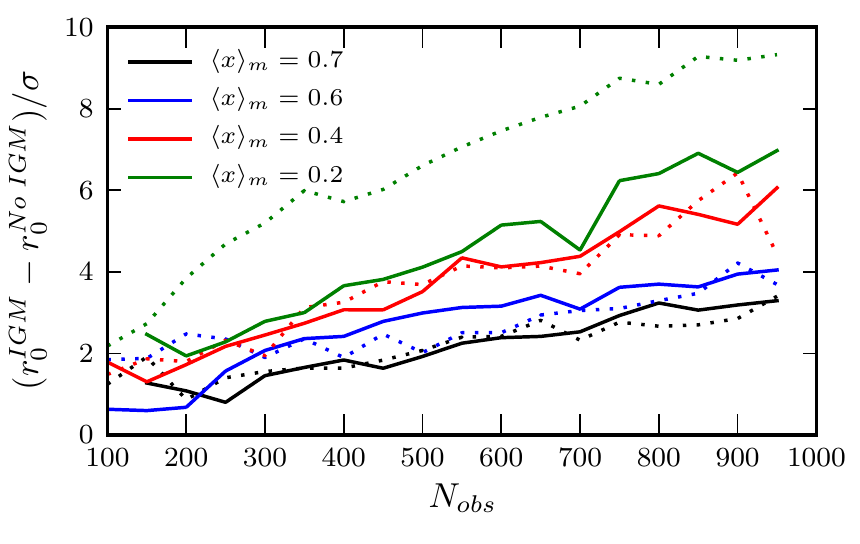}
	\caption{Difference in correlation between the IGM and non-IGM case for different sample sizes and ionized fractions. The solid lines are for the GmG model, and the dotted lines are assuming the single-width Gaussian line model.}
	\label{fig:minobs}
\end{figure}

\begin{figure}
	\centering
	\includegraphics[width=\columnwidth]{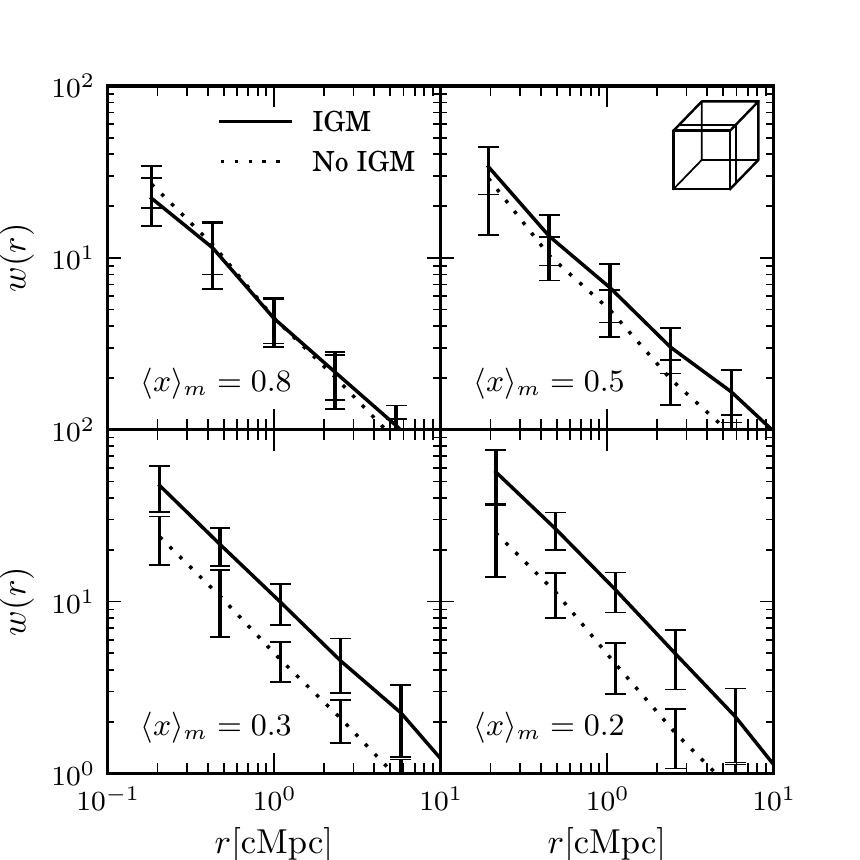}
	\caption{Same as Fig.\ \ref{fig:correlations_wholebox}, but showing the 2D 2PCF for the 500 brightest haloes in the thin box. The error bars in this figure show the $1 \sigma$ field-to-field variance.}
	\label{fig:correlations_thinbox}
\end{figure}

\begin{figure}
	\centering
	\includegraphics[width=\columnwidth]{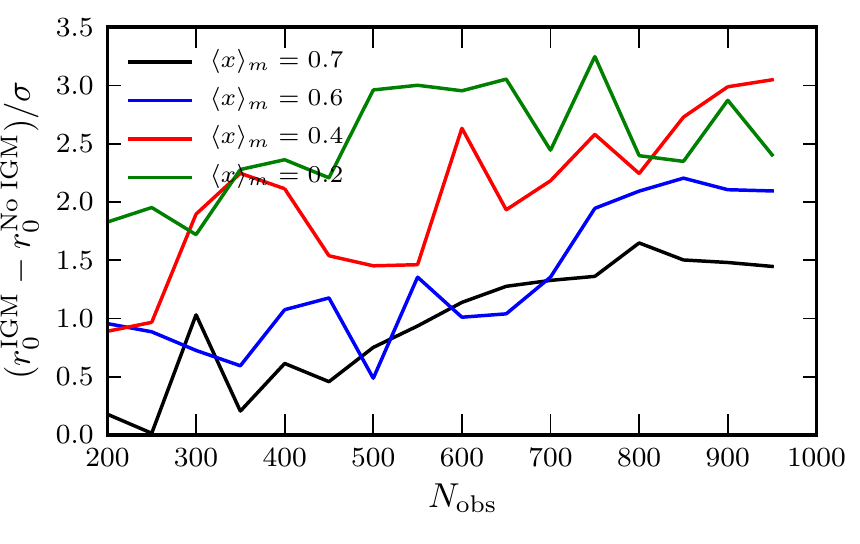}
	\caption{Same as Fig.\ \ref{fig:minobs}, but for the two-dimensional correlation function.}
	\label{fig:minobs_angular}
\end{figure}

In Fig.\ \ref{fig:minobs} we estimate the size of the LAE sample that would be needed to tell a partly neutral IGM apart from a completely ionized one for some different global ionized fractions. This is done by calculating the correlation functions for a number of samples of different sizes and for each sample fitting a power-law of the form:
\begin{equation}
	\xi(r) = \left( \frac{r}{r_0} \right)^{-\gamma}.
	\label{powerlaw}
\end{equation}
For every ionized fraction and sample size, we estimate the uncertainty $\sigma$ in the correlation length $r_0$ for the IGM case using the bootstrap method \citep{efron79}. In Fig.\ \ref{fig:minobs} we show the difference between $r_0$ with and without IGM in terms of $\sigma$. As was hinted at already in Fig.\ \ref{fig:correlations_wholebox}, the correlation method becomes much more effective at lower ionized fractions. For example, at $\xm = 0.2$, already a sample size of 150-200 LAEs would be enough for a $2 \sigma$ detection of IGM-boosted galaxy clustering. However, at $\xm=0.7$ a sample size of up to 500-600 would be needed.

Fig.\ \ref{fig:minobs} appears a bit more pessimistic than for instance \cite{mcquinn2007}, who concluded that a sample size of 250 would be enough to distinguish $\xm = 0.7$ from $\xm=1$. There are two reasons for this. The first is our assumed intrinsic line model. With the Gaussian intrinsic line, used by \cite{mcquinn2007}, the IGM transmission becomes more dependent on the size of the surrounding \hii bubble, as we saw in Sec.\ \ref{sec:t_mass}, which is exactly the effect that causes the correlation to become stronger in a neutral IGM. With the more realistic, double-peaked, line the IGM absorption is more similar across different haloes, especially at higher $\xm$ (see Fig.\ \ref{fig:transmissions}), and the extra clustering is weaker. The second reason is the fact that we have included some random scatter in the mass-to-light relation, which further reduces the clustering signal.

As we saw in Fig.\ \ref{fig:detectability}, to get a sample of several hundreds of LAEs at $\xm = 0.4$ from the deep box, one would need to detect objects down to a detection limit of at least $10^{42}$ erg/s. While obtaining spectra for hundreds of targets is unrealistic with today's ground-based observatories, it may well be possible for a telescope such as the JWST. Using the NIRSpec instrument on the JWST, an exposure time of $\sim$ 30 hours would be enough to obtain \Lya spectra (and thus 3D positions) with reasonable S/N for sources down to $10^{42}$ erg/s (E. Zackrisson, private communication), although uncertainties in the knowledge of the intrinsic line shapes may introduce complications in the translation of the spectra to source redshifts.

In Fig.\ \ref{fig:correlations_thinbox} we show the 2D correlation functions (see Eq.\ \eqref{eq:angular_correlation}; note however that we are showing the projected distance rather than the angle on the sky), calculated with the Landy-Szalay estimator for our thin sub-box. Judging by this figure, it is not surprising that Ou10 do not see any effects of IGM-amplified clustering in their sample of 207 LAEs at $z=6.5$. Even with a sample size of 500, we need about 70\% neutral fraction to reliably distinguish the IGM case from the non-IGM case. 

This is quantified further in Fig.\ \ref{fig:minobs_angular}, which shows the same thing as Fig.\ \ref{fig:minobs}, but for the 2D 2PCF. As expected, a bit of information is lost when throwing away one of the coordinates. Nevertheless, if the IGM is only 40 \% ionized at $z=7$, a sample of ~300 LAEs might very well be enough to detect the IGM clustering at this redshift. Fig.\ \ref{fig:detectability} suggests that a detection limit of $10^{42}$ erg/s would be sufficient, which is only a factor 2.5 lower than the detection limit of Ou10 at $z=6.5$. In fact, the right-most parts of Figs.\ \ref{fig:minobs} and \ref{fig:minobs_angular} may be somewhat conservative. Since we only consider haloes down to $10^{10} \Msol$, the largest samples will have an artificial lower-mass cutoff which makes the IGM and non-IGM samples more similar than if even lower-mass haloes had been included. Thus, the difference in correlation functions could actually be slightly larger than shown here.

\section{Summary and Discussion}
\label{sec:discussion}

\subsection{Summary}
We have presented results from simulations of LAEs during the epoch of reionization and compared these to observations in order to investigate to what extent we may learn something about the global ionized fraction of the IGM at different redshifts. To simulate the observed properties of LAEs, we devised a model that splits the radiative transfer of \Lya photons into two regimes: one (circum-)galactic and one extra-galactic part. For the (circum-)galactic part, we approximate the emerging line shape as a double-peaked Gaussian minus a Gaussian (GmG), while for the extra-galactic part we carry out radiative transfer through our cosmological IGM box. We take the \Lya luminosity to be proportional to halo mass, with a log-normal random scatter, and we fit the mass-to-light ratio to the \Lya luminosity function at $z\sim6$, measured by \cite{ouchi2010}. 

We have shown that a double-peaked model such as our GmG produces some significant differences in results compared to assuming a single Gaussian, as was done in e.g.\ \cite{iliev2008a}, \cite{mcquinn2007} and \cite{dijkstra2007}. In general, our double-peaked model makes the transmitted fraction of \Lya, $T_{\alpha}$, less sensitive to damping wing strength. This is because in this model, most of the emission is offset from the line centre as it enters the IGM, and $\langle T_{\alpha} \rangle$ will be closer to 50 \%, with less dependence on the size of the surrounding \hii bubble.

Comparing to observed LAE LFs at $z=6.5$ by \cite{ouchi2010} and \cite{kashikawa2011}, and assuming that the evolution in the LAE LF is due only to the ionization of the IGM, we have seen that our simulations best match the observations for an IGM ionized fraction of $\xm \sim 0.5$ at $z=6.5$. This is lower than the value of $\langle x \rangle \gtrsim 0.8 \pm 0.2$, claimed by \cite{ouchi2010} (who modelled part of the change in LF as being due to intrinsic LAE evolution) but roughly consistent with the value of $\langle x \rangle \sim 0.62$ given by \cite{kashikawa2011}. We note that both the LFs given in \cite{kashikawa2011} and that in \cite{ouchi2010} give us the same value for $\xm$, despite the fact that these samples are from two fields that are well separated from each other, suggesting that the change is truly global and not due to sample variance.

Furthermore, we compared our simulations to observed equivalent width distributions, and again found that a very neutral IGM is required if the changes between $z\sim6$ and $z\sim 7$ are to be explained by IGM evolution only. Finally, we have given predictions for the sample sizes needed to measure the increase in LAE clustering due to a partly neutral IGM. The two-point correlation function does not require as strong assumptions about the intrinsic evolution of LAEs, and could thus serve as a valuable independent probe of the ionized state of the IGM.

\subsection{How reliable are constraints from LFs and REW distributions?}
Our comparisons to observed LAE luminosity functions and equivalent width distributions seem to be in rough agreement with other studies such as \cite{kashikawa2011} and \cite{pentericci2011} in requiring a very high neutral fraction at $z=6.5$ and $z=7$, but this hinges on the assumption that the evolution in these observables is due to the IGM only and that intrinsic galaxy evolution is negligible. If we set aside the many caveats for a moment and take the values of $\xm = 0.5$ at $z=6.5$ and $\xm = 0.4$ at $z=7$ at face value, this would mean a somewhat later reionization than the scenario we used to produce our IGM boxes (Fig.\ \ref{fig:ionization_hist}). In the reionization scenario used in our simulations, the IGM is ionized only by galaxies, with small sources turned off as their environment becomes sufficiently ionized. By adjusting the assumptions regarding star formation and escape fraction of ionizing photons, it is possible to obtain scenarios in which reionization finished later \citep{iliev2011}.

In the scenario used to produce our IGM boxes, the IGM goes from $\xm \sim 0.4$ to $\xm \sim 1$ between $z=7.7$ and $z=6.5$. This corresponds to a time interval of approximately 170 Myr, which is almost exactly the same as the time interval between $z=7$ and $z=6$ --- the redshift range for the $\xm=0.4$ to $\xm\sim1$ transition implied by our comparisons to observations. So while the particular reionization scenario used here does not perfectly match the observations in terms of \emph{when} reionization took place, it seems that the implied rapid change in ionized fraction is not unreasonable. It may also be objected that a late reionization will result in an electron optical depth too low to match the observations by WMAP \citep{komatsu2009}. However, a fair comparison would require a knowledge of the earlier stages of reionization as well. If, for instance, reionization starts early with small sources there may be time to build up a sufficient electron density before galaxies start rapidly reionizing the IGM at lower redshifts \citep{ahn2010}.

A more serious problem is that these high neutral fractions are incompatible with other observational results. For instance, \cite{raskutti2012} study the IGM temperature in quasar near-zones and find that reionization must have been completed by $z>6.5$ at high confidence, and \cite{hu2010}, \cite{kashikawa2011} and \cite{ouchi2010} study the \Lya line shapes of LAEs at $z=6.5$ and find no evidence of damping wings.

One way to resolve this conflict would be to drop the no-intrinsic-evolution assumption, and allow part of the change in the LAE luminosity functions between $z=6$ and $z=6.5$ to be explained by intrinsic galaxy evolutions, as argued by, for instance, \cite{dayal2008} and \cite{ouchi2010}.
As we noted in Sec.\ \ref{sec:lumfunc_comparison}, we can reproduce the drop in LF by galaxy mass-evolution only, although we then run into problems at lower redshifts. 

A second solution was recently suggested by \cite{bolton2012} 
who showed that dense small-scale \hi structures,
below the resolution of our study, may play an important role in
limiting the observability of LAEs even when the global neutral
fraction is low. However, their results assume a constant UV
background and it is not yet clear how this translates to the more
realistic case of a fluctuating ionizing rate which is higher in the
denser regions close to sources.

In conclusion, it would seem that any constraints on the global IGM ionized fraction obtained from LFs or REW distributions will remain uncertain at best until more is known about the intrinsic evolution of LAEs at $z>6$ and the role and prevalence of small-scale \hi structures. The degeneracy between intrinsic evolution and IGM ionization could possibly be broken with better observations of the UV LF evolution of LAEs \citep{ouchi2010}, or by 21 cm experiments such as LOFAR and SKA.

In addition to the uncertainties related to intrinsic LAE evolution, we have shown in this paper that simplifying assumptions regarding \Lya line shapes can have significant effects on the results of simulations. Line shapes with most of the emission to the sides of the line centre require more neutral hydrogen to obtain a given drop in transmitted fraction. As an example, if we redo our luminosity function calculations shown in Fig.\ \ref{fig:lae_lfs_6p5} using the Gaussian line profile, we get a best fit to the observations for $\xm \sim 0.7$ as compared to $0.5$ with the GmG model. While few spectroscopic observations of sufficient resolution exist at high redshift, lower-redshift observations (e.g.\ \citealt{tapken2007,verhamme2008,yamada2012}) show that most lines are either double-peaked or with a single peak that is offset from the line centre. For our purposes, these offset single-peaked lines are equivalent to the double-peaked ones --- the important distinction is between the Gaussian lines with the emission centred at the line centre, and those where it is offset from the line centre. While we believe our GmG model to be more realistic that a single Gaussian, lower-$z$ observations do show that most LAEs have \emph{some} emission at the line centre (e.g.\ \citealt{tapken2007}), even when a large part of the emission is shifted away from the centre. If this is true also at higher $z$, it would mean that the GmG model (which has zero emission at the line centre) tends to slightly under-predict the effect of a neutral IGM. 

\subsection{Future prospects}

While both the luminosity functions and equivalent width distributions suffer from degeneracies and model-dependencies, the two-point correlation function could be used to put some of the model assumptions to the test. After correctly plotting the results from \cite{iliev2008a}, it seems all theoretical models now agree that a significant neutral fraction should give an appreciable boost to the correlation function. While our line model and the fact that we allow some random scatter in the mass-to-light ratio make our predictions for the two-point correlation function method as a means for constraining the IGM ionized fraction somewhat more pessimistic than \cite{mcquinn2007}, it still seems that the method could be useful once observed LAE samples grow by a factor of two or three. Our analysis here has been focused only on the amplification of the clustering induced by the patchiness of the reionization process. Other effects such as intrinsic LAE bias \citep{orsi2008} and environment dependent selection effects \citep{zheng2011} may cause LAEs to become more clustered than drop-out selected galaxies, but it is hard to imagine that these effects would change rapidly with redshift.

As we show in Fig.\ \ref{fig:minobs_angular}, if only 2D positions are available, a sample size of many hundreds of LAEs is required for any meaningful constraints on $\xm$. It is thus not very surprising that \cite{ouchi2010} do not measure any IGM-induced boost in clustering with their sample of 207 galaxies. Using our model, their non-detection gives only a weak lower limit of $\xm \gtrsim 0.4$. With future, larger, samples it may be possible to put more meaningful constraints on $\xm$ using LAE clustering. With a sample $\sim$ 700 LAEs we would expect a 2 $\sigma$ detection of IGM-amplified clustering if the IGM is $\gtrsim 50$ \% neutral, which could be used to test the predictions from luminosity functions and equivalent width distributions assuming no intrinsic galaxy evolution. With 1000 LAEs, even a neutral fraction of 30 \% should give a strong clustering signal. With the new Hyper Suprime-Cam at the Subaru telescope, it is expected that future surveys will detect up to 10 000 LAEs at $z=6.5$ over a 30 deg$^2$ area on the sky (M. Ouchi, private communication). A detailed theoretical investigation of such a survey would require much larger-scale simulations than the ones presented here, but simple extrapolation of Fig.\ \ref{fig:minobs_angular} would suggest that 10 000 detected LAEs can put strong constraints on the global ionized fraction.


\section*{Acknowledgements}

This study was supported in part by the Villum Foundation
and the Swedish Research Council grant 2009-4088.
The authors acknowledge the Swedish National Infrastructure for Computing (SNIC) resources at HPC2N (Ume\aa, Sweden) and PDC (Stockholm, Sweden), TeraGrid/XSED{E} and
the Texas Advanced Computing Center (TACC) at The University of Texas at
Austin (URL: http://www.tacc.utexas.edu) for providing HPC resources.
The SPH simulations and the galactic \Lya RT were performed at the facilities provided by the Danish Center for Scientific Computing.
The authors also acknowledge the Royal Society International Joint Project grant.
ITI was supported by The Southeast Physics Network (SEPNet) and the Science and
Technology Facilities Council grants ST/F002858/1 and ST/I000976/1.
PRS was supported by  NSF grants AST-0708176 and AST-1009799, NASA grants NNX07AH09G,
NNG04G177G and NNX11AE09G, and Chandra grant SAO TM8-9009X.
We are very grateful to Masami Ouchi for useful comments and information about the upcoming Hyper Suprime-Cam survey.
We would also like to thank Anne Verhamme for useful discussions about resolution effects when simulating \Lya spectra, and Erik Zackrisson for providing the calculations on JWST detection limits.

\bibliographystyle{mn} \bibliography{refs}

\label{lastpage}
\end{document}